\documentclass[11pt,a4paper]{article}

\usepackage{graphicx}

\usepackage{amsmath,amssymb}

\usepackage{jinstpub}
\usepackage{siunitx}[=v2]
\DeclareSIUnit\ppm{ppm}
\usepackage{booktabs}
\usepackage[backgroundcolor=yellow!15, bordercolor=orange]{todonotes}
\usepackage[draft,inline=true]{fixme}
\usepackage[noabbrev]{cleveref}
\usepackage{pdflscape}
\usepackage[shortcuts]{extdash}
\usepackage{lineno,hyperref}

\begin{document}
\title{Wideband precision stabilization\\of the -18.6$\,$kV retarding voltage\\for the KATRIN spectrometer}
\emailAdd{rodenbeck@wwu.de}
\author[a]{C. Rodenbeck,}
\author[b]{S. W\"ustling,}
\author[c]{S. Enomoto,}
\author[b]{J. Hartmann,}
\author[a]{O. Rest,}
\author[d]{T. Th\"ummler,}
\author[a]{and C. Weinheimer}

\affiliation[a]{Institut für Kernphysik\\
Westfälische Wilhelms-Universität Münster\\ Münster, Germany.}
\affiliation[b]{Institute for Data Processing and Electronics (IPE)\\
Karlsruhe Institute of Technology (KIT)\\
Eggenstein-Leopoldshafen, Germany}
\affiliation[c]{Center for Experimental Nuclear Physics and Astrophysics, and Department
of Physics\\University of Washington\\ Seattle, WA, USA.}
\affiliation[d]{Institute for Astroparticle Physics (IAP)\\
Karlsruhe Institute of Technology (KIT)\\
 Eggenstein-Leopoldshafen, Germany.}

\date{\today}

\maketitle

\begin{abstract}

The Karlsruhe Tritium Neutrino Experiment (KATRIN) measures the effective
electron anti-neutrino mass with an unprecedented design sensitivity of
\SI{0.2}{\electronvolt} (\SI{90}{\percent} C.L.). In this experiment, the energy
spectrum of beta electrons near the tritium decay endpoint is analyzed with a
highly accurate spectrometer. To reach the KATRIN sensitivity target, the
retarding voltage of this spectrometer must be stable to the ppm (\num{1e-6})
level and well known on various time scales ($\mu s$ up to months), for values
around \SI{-18.6}{\kilo\volt}. A custom-designed high-voltage regulation system
mitigates the impact of interference sources in the absence of a closed electric
shield around the large spectrometer vessel. In this article, we describe the
regulation system and its integration into the KATRIN setup. Independent
monitoring methods demonstrate a stability within \SI{2}{\ppm}, exceeding
KATRIN's specifications.

\end{abstract}

\newpage

\section{Introduction}\label{sec:intro}

The Karlsruhe Tritium Neutrino (KATRIN) \cite{DesignReport05, DesignReport21}
experiment is measuring the effective mass of the electron antineutrino with a
targeted sensitivity of $0.2\,\mathrm{eV/c^2}$ (at $90\,\%$ confidence level)
within three live years of data taking. KATRIN examines the kinematics of
tritium beta decays in a model-independent approach. The tritium beta-decay
electron spectrum is measured near the endpoint at $18.6\,\mathrm{keV}$, where
the effect of the neutrino mass is the largest.

The experiment consists of a Windowless Gaseous Tritium Source (WGTS), a
transport and pumping section \cite{Marsteller2020}, two spectrometers of MAC-E
filter (Magnetic Adiabatic Collimation with an Electrostatic filter) type
\cite{Beamson1980,Lobashev1985,Picard1992}, and a focal plane detector
\cite{amsbaugh2015}. The differential and cryogenic pumping sections prevent
tritium gas inside the WGTS from entering the spectrometer. The beta-decay
electrons are magnetically guided to the spectrometers. A negative high voltage
is applied to the main spectrometer vessel, creating a retarding potential for
the electrons entering the spectrometer. Only electrons above a threshold energy
can penetrate the potential to reach the detector. An energy spectrum of the
electrons is recorded at the detector. The value of the retarding potential
defines the energy scale for every electron spectrum.

Any high-voltage instability introduces a bias on the neutrino-mass
measurement. To achieve KATRIN's physics goal, the standard deviation of any
instability of the energy scale should be below $60\,$meV, corresponding to a
stability requirement of $3\,$ppm (\num{3e-6}) at \SI{-18.6}{\kilo\volt}
on a wide range of time scales ($\mu s$ up to months).

A measurement chain with two precision high-voltage dividers~\cite{K3509,K6513}
has been built and commissioned. It is combined with a custom-made,
MHz-bandwidth regulation loop, called ``post-regulation.'' The post-regulation
stabilizes and sets the retarding voltage with the required precision.

The setup, its recent upgrade, and the performance of the post-regulation is
presented in this paper.

\section{Retarding high voltage at the KATRIN main spectrometer}\label{sec:generalSetup}

The spectrometer is a large, barrel-shaped steel vacuum vessel with a maximum
diameter of \SI{9.8}{\meter} in its cylindrical section and a total length of
\SI{23.28}{\meter} \cite{DesignReport21}. Inside the main spectrometer, an
electric field is created by a high voltage applied directly to the vessel
itself. The electric field is fine-tuned with an inner electrode system
consisting of full metal cones at the entrance and exit and two layers of wire
grid electrodes in between. The inner electrode system is electrically biased by
about \SI{-100}{\volt} (adjustable to the requirement of the measurements)
relative to the spectrometer vessel (cf. \cite{DesignReport21}, pp. 55 ff.). In
the nominal symmetric configuration the maximum retarding potential $qU$ for
electrons of charge $q=-\mathrm{e}$ and the minimum electric field are in the
central plane of the spectrometer. A superconducting magnet at each ends of the
vessel and an air coil system create a guiding magnetic field. In the nominal
configuration, the magnetic field has its minimum ($\approx \SI{1e-4}{T}$) at
the central plane of the spectrometer and its maximum ($\SI{4.2}{T}$) at the
entrance and exit of the spectrometer (cf. \cite{DesignReport21}, pp. 66
ff.). The central plane, where the magnetic and electric field have their
minimum, is called analyzing plane.

\begin{figure}
  \centering \includegraphics[width=0.9\textwidth]{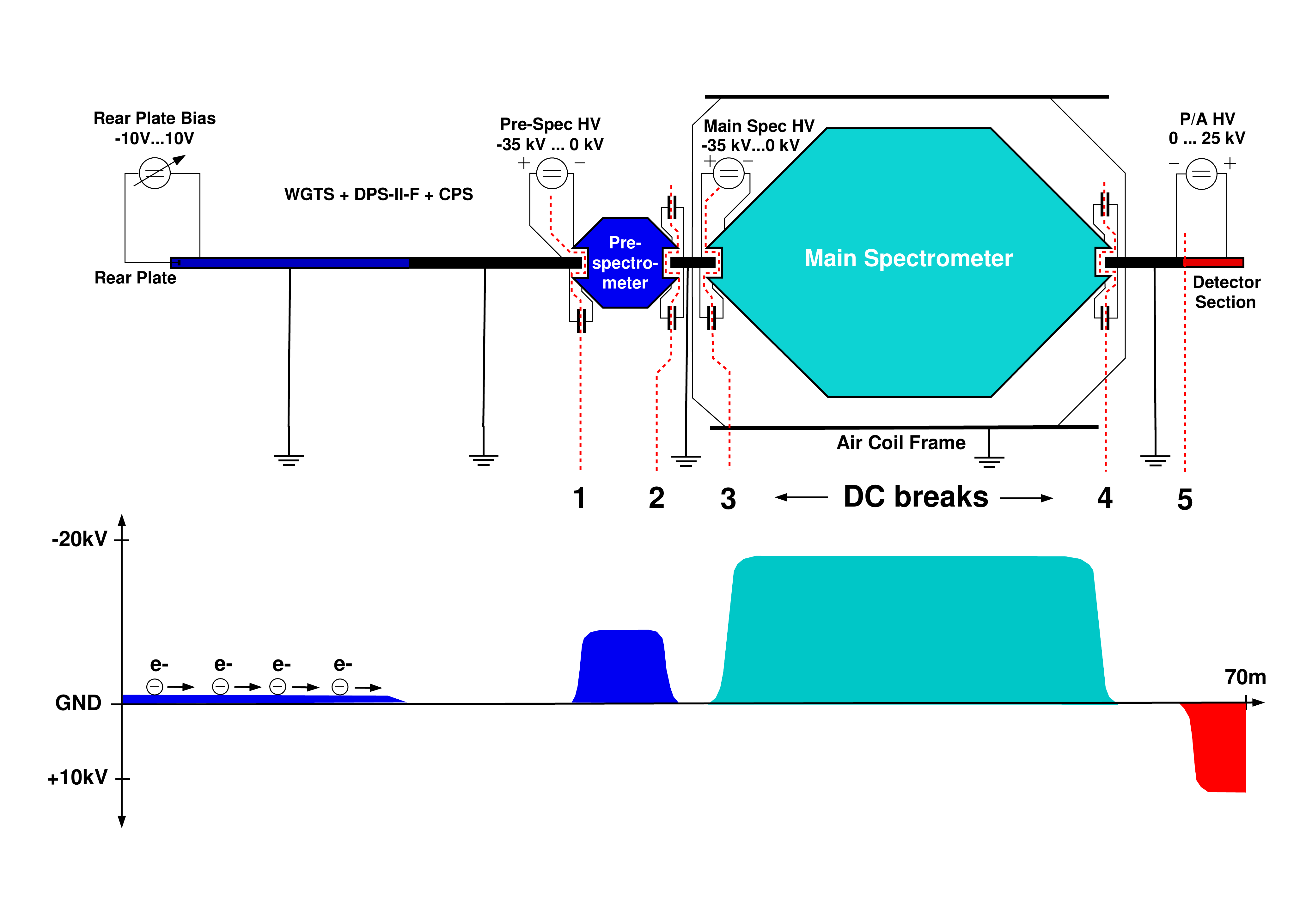}
  \caption{\textbf{Overall electrostatic potential scheme of the KATRIN
      experiment.} The upper plot shows the components of the KATRIN beamline,
    the lower one their electrostatic potential. On the left is the rear wall
    with an adjustable bias voltage. The bias voltage is used to optimize the
    plasma conditions inside the WGTS \cite{KNM22021}. The plasma defines the
    starting potential of beta electrons in tritium source (WGTS). The WGTS and
    the transport and pumping section (DPS + CPS) are grounded. Both
    spectrometers can be set to different retarding potentials. The lower plot
    shows the nominal negative potential settings for the pre- and main
    spectrometer. Located on the right, after the \SI{70}{\meter} beamline is
    the detector with a positive post-acceleration voltage. The red dashed lines
    mark the DC breaks. The main spectrometer potential shown in turquoise is
    the focus of this paper.}
  \label{fig:potential_scheme}
\end{figure}

The transmission condition for beta electrons through the KATRIN main
spectrometer depends on the magnetic field and electric potential settings and
the potential of the WGTS tube, where the electrons are generated by tritium
beta decay. The WGTS tube and the beam tube guiding the electrons to the main
spectrometer are connected to the earth ground and serve as reference potential
(c.f. \cref{fig:potential_scheme}).

Electrons entering the spectrometer experience the Lorentz force by the electric
and magnetic fields inside the spectrometer. In the adiabatic approximation, the
magnetic field gradient keeps the orbital magnetic moment
\begin{equation}
    \mu = \frac{E_\perp}{B}
\end{equation}
constant (equation given in non-relativistic notation). Thus the magnetic field
gradient collimates the electron momentum to be parallel to the magnetic field
line in the center of the spectrometer, transforming the perpendicular or
cyclotron component $E_\perp$ into the longitudinal component $E_\parallel$ of
the kinetic energy. The electric field (cf. \cref{fig:EfieldToF} (a)) transforms
the longitudinal component of the electrons' kinetic energy to potential energy
on their way through the spectrometer. This is the MAC-E filter principle.

From the spectrometer entrance up to the analyzing plane of the spectrometer,
the kinetic energy $E_{\mathrm{kin}}$ of the electrons is decreased. Their
kinetic energy is increased again after the analyzing plane up to the
spectrometer exit, if their kinetic energy is large enough to overcome the
maximum retarding potential, otherwise they are reflected. Large enough kinetic
energy means $E_{\mathrm{kin}} > q U_0$, for electrons with their momentum
already parallel to the magnetiv field lines before entering the spectrometer
(angle $\theta=0$), with $U_0$ the retarding potential and $q$ the charge of the
electron. Therefore, the transmission function of electrons with $\theta=0$ is a
step from no transmission to full transmission at the threshold energy
$E_{\mathrm{kin}}=q U_0$. For electrons with $\theta \neq 0$ the threshold
energy is shifted towards higher values. These electrons need additional surplus
energy to overcome the filter potential.

To quantify the effect on the transmission condition by a time-dependent noise
on $U_0$, dedicated simulations were performed. In the following, the simulation
results are used to define the requirements for the high-voltage
stability. Possible sources for instabilities on different time
scales are discussed.

\subsection{Requirements}\label{sec:requirements}

To explain the general influence of high-voltage noise on electrons flying
through the main spectrometer, a simplified model with on-axis electrons, and
only their guiding motion is chosen. On-axis meaning that the electrons are
following a field line along the spectrometer axis $z$ ($x=y=0$). We consider
the electric potential $U(z)$ and the electric field $E(z)$ as the only
$z$-dependent quantities.

\begin{figure}
  \centering
  \includegraphics[width=\textwidth]{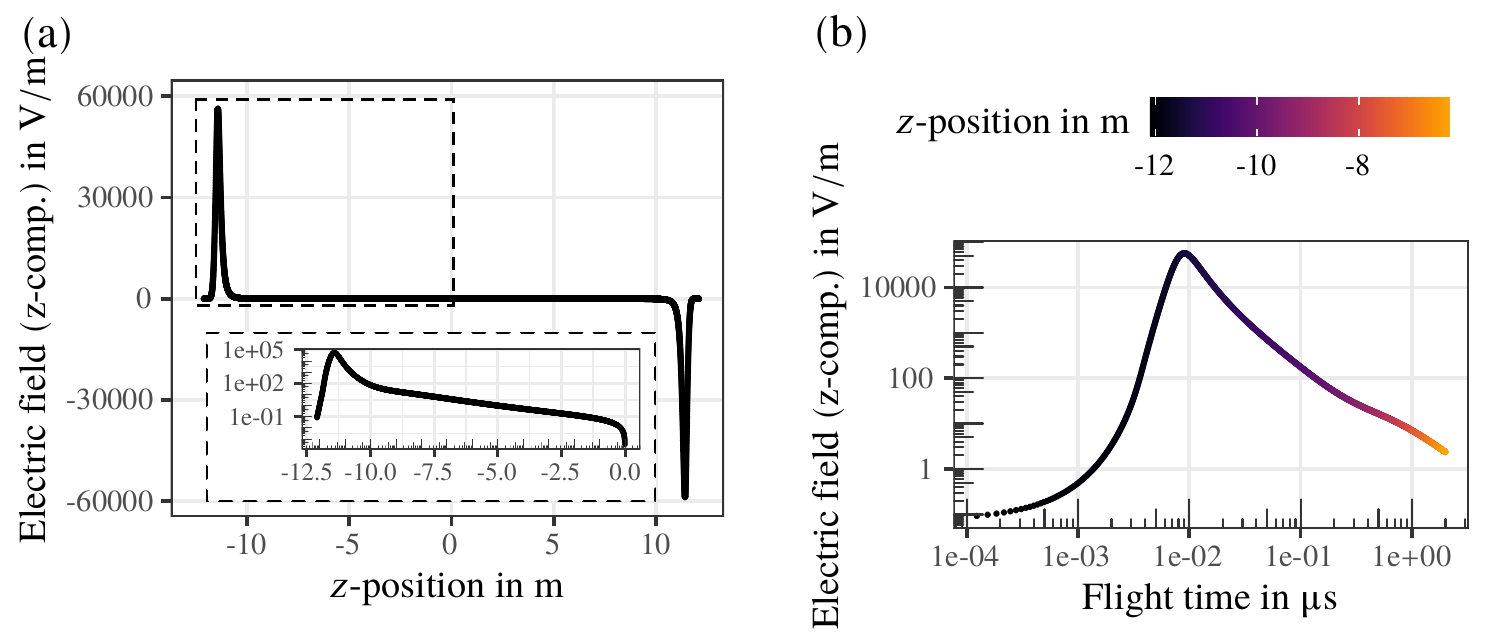}
  \caption{\textbf{Electric field.} Plot (a) shows the z component of the
    electric field along the beam axis $z$ inside the main spectrometer ($z=0$
    center of the spectrometer). The retarding potential of $U_0 =
    \SI{-18500}{\volt}$ matches the standard measurement window below the
    tritium endpoint. Plot (b) shows the z component of the electric field
    plotted against the electron's flight time from the spectrometer entrance to
    a position along $z$. The flight time is calculated for an electron with a
    surplus energy of \SI{0.5}{\electronvolt} and $\theta=0$.}
  \label{fig:EfieldToF}
\end{figure}

The potential inside the spectrometer along the beam axis $z$ can be described
as $U(z) = a(z) \cdot U_0$. The $z$-dependence of the potential seen by the
electron $a(z)$ is defined by the geometry of the spectrometer and can be
determined with the Kassiopeia simulation framework \cite{kassiopeia17} . With a
time dependent fluctuating potential $\Delta U(t)$, the potential becomes:
\begin{align}
  U(z,t)= a(z) \cdot \left( U_0 + \Delta U(t)\right) = U(z) + a(z) \cdot \Delta U(t)
  \cdot \frac{U_0}{U_0} = U(z) \cdot \left(1 + \frac{\Delta U(t)}{U_0} \right).
\end{align}
The electric field is:
\begin{align}\label{eq:electric_field_from_noise}
  E(z) = - \frac{\partial U(t,z)}{\partial z} 
             = -\frac{dU}{dz} \left( 1 + \frac{\Delta U(t)}{U_0} \right).
\end{align}
Since the electric field (c.f. \cref{fig:EfieldToF} (a)) is close to zero at the
analyzing plane, it is virtually a Faraday cage. Here, any time-dependent
variation, with $\Delta U(t) << U_0$ has only a small effect.

To investigate the effect of different noise waveforms, time-of-flight
simulations were performed, building on previous work
\cite{Steinbrink2013,Fulst2020}. In the simulation, the electrons are tracked in
small steps along the $z$-axis inside the spectrometer. At each step their
flight time and momentum are determined. 

In \cref{fig:EfieldToF} (b), the electric field strength is plotted against the flight
time of an electron through the spectrometer. Only the spectrometer entrance,
where the electric field strength is largest is shown. The electron's travel time from
the $z$-position, where the electric field starts to increase
($>\SI[per-mode=symbol]{1}{\volt\per\meter}$) to the $z$-position, where the
electric field starts decreasing again to values below
$\SI[per-mode=symbol]{100}{\volt\per\meter}$, called $t_{\mathrm{trav}}$, is
within \SIrange{1e-3}{0.1}{\micro\second}.

A sinusoidal noise $f_{\sin}(f,t) = A \sin(2\pi f t)$ on the retarding potential
with a fixed amplitude $A$ is used as test noise in the simulations. For
frequency dependency investigations $f$ was changed within
\SIrange{1}{1e9}{\hertz}. For simplification only electrons with fixed angle
$\theta=0$ are considered here.

\begin{figure}
  \centering
  \includegraphics{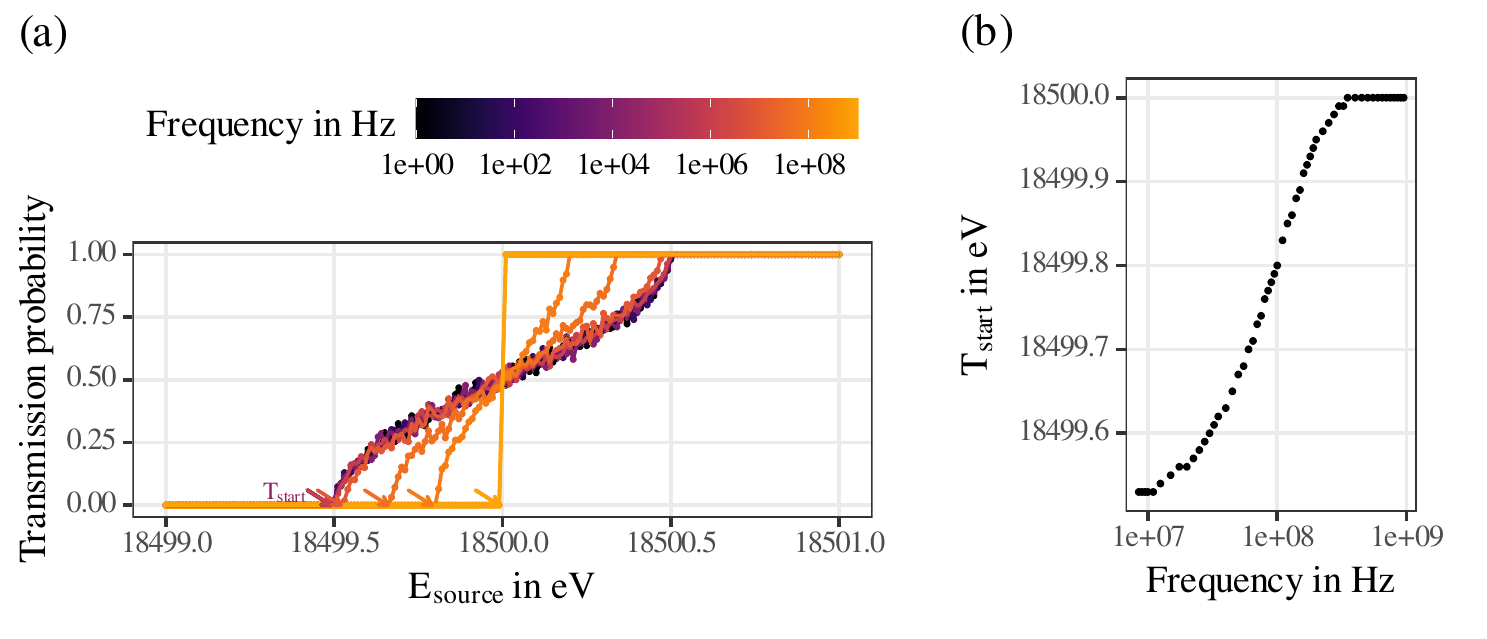}
  \caption{\textbf{Transmission conditions.} The left plot (a) shows the
    transmission function for electrons with $\theta = 0$ and different kinetic
    energies $E_{\mathrm{source}}$ at the spectrometer entrance. The retarding
    potential is fixed to $U_0 = \SI{-18500}{\volt}$ with a sinusoidal
    noise $f_{\sin}(f,t)$ with an amplitude of \SI{0.5}{\volt} and at various
    frequencies $f$. For each frequency \num{100000} electrons are
    generated. The start times are randomized with $t_\mathrm{start} \in
    [0,1/f]$. The right plot (b) shows the minimal $E_{\mathrm{source}}$ for the
    transmission of the electron $T_{\mathrm{start}}$ (indicated with arrows in
    the left plot), as a function of the frequency.}
  \label{fig:TFSimuSinus}
\end{figure}

The result is shown in the left plot of \cref{fig:TFSimuSinus}. The noise
broadens the transmission function by the height of the amplitude $A$,
independent of the frequency for frequencies below \SI{10}{\mega\hertz}. The
independence of the frequency can be explained by $t_{\mathrm{trav}}$. The
change in amplitude of the noise $f_{\sin}(f,t)$ is slow, compared to
$t_{\mathrm{trav}}$. Only for frequencies above
$\SI{10}{\mega\hertz}=\SI{0.1}{\micro\second}$, the noise changes the electric
field within $t_{\mathrm{trav}}$. Therefore for frequencies below
\SI{10}{\mega\hertz}, the broadening effect of the sinusoidal test noise is the
density distribution of $f_{\sin}(f,t)$, without any frequency dependency. For
frequencies above \SI{10}{\mega\hertz} the broadening effect on the transmission
function starts to attenuate and vanishes for $f >\SI{400}{\mega\hertz}$, see
right plot in \cref{fig:TFSimuSinus}.

The broadening of the transmission function by a noise on the retarding
potential directly shows the importance to eliminate its impact as much as
possible, and to measure the residual broadening. Any unknown broadening of the
energy scale, defined by the transmission function, leads to a shift of measured
the neutrino mass by \cite{Robertson1988}
\begin{align}
  \delta m^2_{\nu} = -2 \sigma^2\,. \label{eq:broad}
\end{align}
In the KATRIN design report \cite{DesignReport05}, the uncertainty budget was
specified to reach the design sensitivity. A neutrino mass shift of
\SI{7.5e-3}{\electronvolt\square} was specified, leading to a maximum broadening
of the energy scale of
\begin{equation}\label{eq:noiseLimit}
    \sigma=\SI{60}{\milli\electronvolt}\,.
\end{equation}
This is equivalent to an unknown Gaussian broadening of the retarding potential
with a maximal standard deviation of \SI{60}{\milli\volt} on all time scales:
from the length of one measurement campaign (roughly 10 weeks
\cite{DesignReport21}), down to \SI{0.1}{\micro\second} (\SI{10}{\mega\hertz}).

\subsection{Short-term precision in a noisy environment}

Apart from residual voltage ripple that may originate from the high voltage
supply, the stability of the retarding voltage of the KATRIN main spectrometer
can be compromised by external interference through electromagnetic coupling. An
ideal countermeasure against such interference would be to place the
spectrometer vessel in an overall electric shielding i.e. a Faraday
cage. However, this would be impractical for the KATRIN apparatus. The strong
capacitive coupling of the inner electrodes and the tank vessel prevents its use
as a high-frequency shielding for the inner electrodes in view of the low noise
requirement \cref{eq:noiseLimit}. As a consequence, any alternating
electromagnetic fields in the vicinity of the spectrometer vessel represent
possible interference sources for the retarding voltage. Alternating stray
electric fields arise, for example, from AC power lines or radio
transmitters. Moreover, there is an own floating 3-phase \SI{400}{\volt} AC
power network attached to the vessel, which is fed by a large insulation
transformer with a $\num{3}\times\SI{6}{\kilo\volt\ampere}$ rating. This
floating AC supply is required mainly for the vacuum pumps
(cf. \cite{DesignReport21}, pp. 50 ff.) but also for various data acquisition
and monitoring purposes. Although this transformer employs primary and secondary
shield windings, there is a remaining capacitive coupling bringing about certain
mains voltage interferences. Also, the large-area grounding network is sensitive
to electromagnetic resonant pickups with possible effects on the retarding
voltage. To achieve short-term precision of the retarding voltage, excessive
high-frequency components need to be absent. As soon as high-frequency signals
present in geometrically large apparatus are discussed, it is essential to
specify \textit{at which position} on the apparatus the high-frequency voltage
component is most detrimental. Concerning the KATRIN main spectrometer,
providing a smooth retarding potential within the spectrometer volume can be
established by restoring a Faraday cage configuration of the beam tube, while DC
voltage breakpoints (isolators) are present at the spectrometer entrances and
exits. Especially at the electrical isolator, located at the spectrometer
entrance, any noise must be reduced below a tolerable value
(cf. \cref{sec:requirements}).

\subsection{Long-term precision and absolute accuracy}\label{sec:measChain}

A precision measurement chain is in place \cite{DesignReport05} to cover the
requirements on the longer time scales. It consists of two custom-built
precision high voltage dividers K35 \cite{K3509} and K65 \cite{K6513} and
\num{8.5}-digit precision digital voltmeters\footnote{Fluke
8508A}. During measurements at the KATRIN beamline, one of the dividers is
connected to the inner electrode system at the electrode providing the highest
retarding energy $qU_0$ and thus defining the analyzing plane of the
spectrometer. The 1972:1 tap of the K35 leads to a reading in the \SI{10}{\volt}
range when operated at the nominal KATRIN standard measurement window around
$U_0=\SI{-18500}{\volt}$. The same is true for the K65 divider, which provides a
1818:1 tap.

Voltage measurements with sub-ppm accuracy \cite{Iso5727} in the
range of $\pm\SI{20}{\volt}$ can be achieved by using a voltmeter with high
precision and regular calibrations to achieve trueness. The voltmeters are
calibrated two times per week during measurement campaigns with multiple
\SI{10}{\volt} reference standards\footnote{Fluke 732B}. A subset of the
reference standards is calibrated annually using a Josephson standard
\cite{DesignReport21}. During regular measurements, the voltmeter readout rate is
at \num{0.5} samples per second.

The dividers are calibrated regularly with independent methods. One method is
an absolute calibration method \cite{AbsCal19}, while the others involve
spectroscopic measurements of mono-energetic conversion electrons from
$^{\mathrm{83m}}$Kr electron capture decays \cite{kryptonHvCal2017}. Both
methods are performed on-site and give matching results. They are also in
agreement with cross calibration measurements at the Physikalisch-Technische
Bundesanstalt (PTB) \cite{K3509, K6513}. The calibration history of both
dividers, covering more than ten years, shows an excellent long-term stability
to better than the ppm-level over a year \cite{kryptonHvCal2017}.

This unique combination of precision high voltage dividers and voltmeters with
regular calibrations and independent calibration methods achieves a robust
measurement of the retarding potential on the sub-ppm level on time scales from
\SI{2}{\second} up to years.

\section{Motivation for a custom-designed high voltage system}

As outlined in \cref{sec:requirements}, the requirements for retarding voltage
not only imply an ultra-high stability over long time scales but also a low
ripple up to frequencies in the MHz range. At the same time, the high-voltage
generation system has to cope with a multitude of interference sources. To
cover the long-term requirements, world-leading DC voltage measurement
technology is being employed as shown in \cref{sec:measChain}. The design and
implementation of the high-voltage generating system involved investigations of
the suitability of commercially available power supplies to meet KATRIN's
stringent technical and scientific requirements. Typically, commercial power
supplies for the voltage range of interest (up to \SI{-35}{\kilo\volt}) and the
required power (several \si{\milli\ampere} to drive voltage dividers, etc.) are
switched-mode devices that operate at switching frequencies of some tens of
\si{\kilo\hertz}. They are available with ppm DC stability. However, the
voltage regulation behaviour of these switched-mode supplies is quite slow. A
principal limit to their regulation speed is given by the switched-mode
principle. In practice, the regulation loop has to be even slower due to the
loop stability requirements considering the slow speed of their multiplier
cascade. There is a gap between what precision high-voltage DC metrology in
combination with a commercial high voltage power supply can do and the high
voltage stability requirements of the KATRIN experiment. Bridging this gap calls
for a high-voltage system that allows low high-voltage source impedance for
frequencies up to the MHz range.

Another important requirement for the high-voltage supply system to be designed
is its output range. For neutrino mass measurements the retarding potential is
varied around the endpoint at \SI{-18.6}{\kilo\volt} in different step sizes
from \SI{1}{\volt} up to \SI{200}{\volt} \cite{KNM1Ana21}. For systematic
measurements the retarding potential is also needed at other values; for example
the krypton conversion electron spectroscopy of $^{\mathrm{83m}}$Kr
\cite{krypton2017} requires the range of \SIrange{-32}{-7}{\kilo\volt}. Not only
is that voltage range different but also the scanning strategy. For
$^{\mathrm{83m}}$Kr spectroscopy the voltage is changed often (about every
\SIrange{20}{180}{\second}) in very small steps of \SI{30}{\milli\volt} up to
steps of \SI{10}{\volt}, whereas the change frequency for tritium measurements
is less frequently (varying roughly between \SI{30}{\second} to
\SI{20}{\minute}) in larger steps of around \SI{2}{\volt} up to
\SI{300}{\volt}. Any voltage changes require additional settling times. To make
measurements as efficient as possible the settling time needs to be minimized.

\section{Implementation of the high voltage system}
Typically one would use electric capacitors, to create a low impedance for AC
voltages, while at the same time maintaining isolation for a high DC voltage.
However, at the required low impedance values and capacitor voltage ratings,
passive capacitive filtering is only viable above about 500kHz. Considering
frequencies such as the 50 Hz power grid frequency, unrealistically large
capacitors, storing large amounts of energy and being a safety hazard, would be
needed to reduce the overall interference level of the KATRIN main spectrometer.

Another method to obtain the same effect is to use an active voltage
regulator. It consists of a precise pickup of the voltage present between the
respective electric nodes, an adjustable precision voltage source for the
setpoint, an error amplifier, a PID regulator, and an actuator that can control
the voltage between the nodes. If a fast regulator action is required, all these
components must be reasonably fast and exhibit low time delays. As outlined
earlier, commercial HV power supply units do contain a voltage regulator, but it
does not cover the required bandwidth into the MHz range.

The solution is a nested regulation approach. The primary high-voltage
generation is provided by a commercial switched-mode high voltage power supply,
which provides a good efficiency over a wide output voltage range. A second,
linear regulation loop with a control range of only a few volts provides speed
and fine regulation. Pick-up of the actual voltage is performed directly at the
load for a maximum bandwidth. In our case, pick-up of the momentary retarding
voltage is performed directly across the beam tube insulator at the main
spectrometer entrance. To provide a low AC impedance even above the
regulator upper cutoff frequency, ceramic filtering capacitors are arranged
also at the beam tube insulator at the spectrometer entrance.

\subsection{Philosophy of the regulator structure}\label{sec:regulatorStruct}

\begin{figure}[t!]
  \centering
  \includegraphics{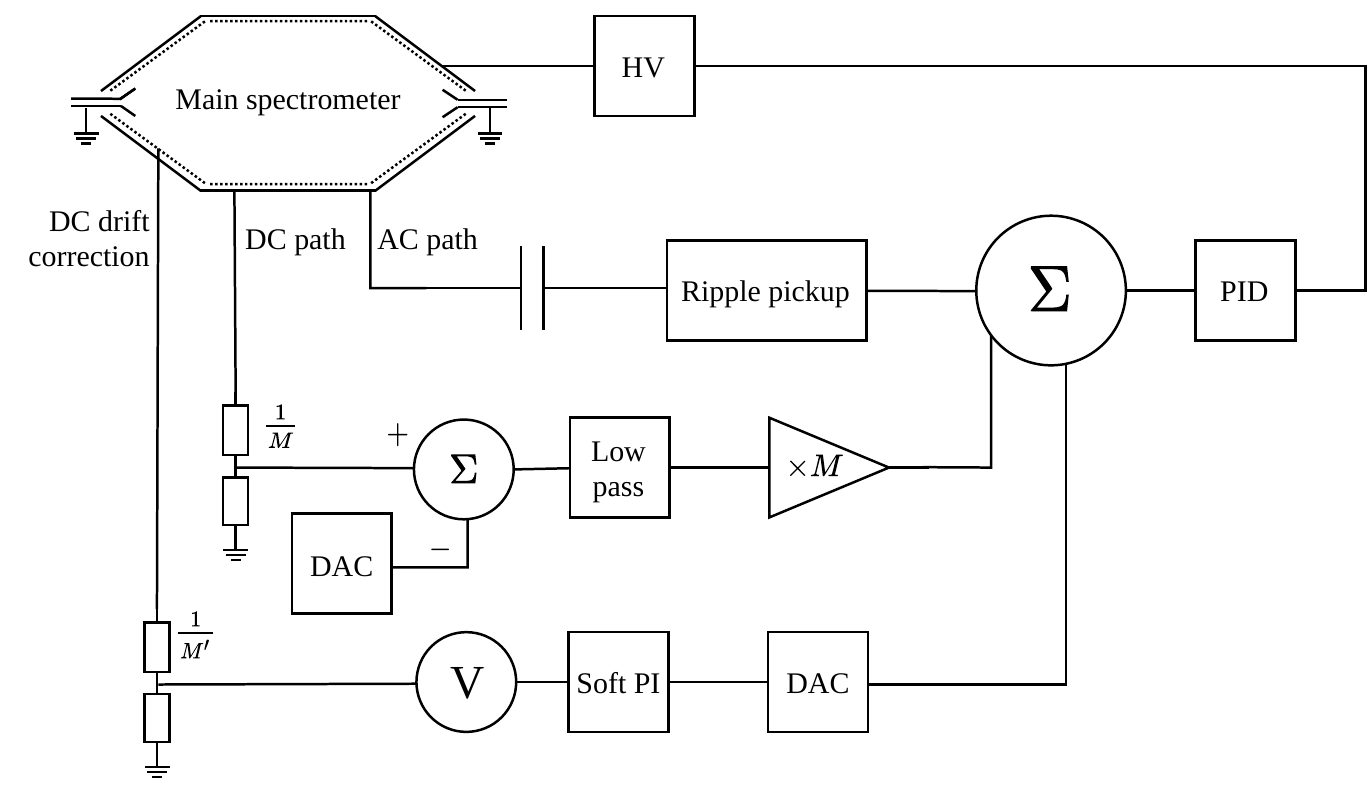}
  \caption{\textbf{Schematic diagram of the regulator structure}. The schematic
    shows the nested approach with the three paths: AC path, DC path, and the
    outer loop with the DC drift correction. The dashed line in the main
    spectrometer visualizes the inner electrode system (cf. \cref{sec:generalSetup}). A more detailed block diagram is shown in \cref{fig:blockDiagrammPR}.}
  \label{fig:pr-schematic}
\end{figure}

To obtain a high-precision high-bandwidth readout and regulation of the actual
voltage, a three-path scheme is employed. Smoothing is obtained by an AC-coupled
control loop while DC regulation is obtained by a DC-coupled loop. The primary
DC loop does not need to fulfill the full long-term high voltage stability
requirements because the precision voltage divider along with its voltmeter
readout (cf. \cref{sec:measChain}) can be integrated into the system by a
secondary, software-based DC drift correction loop (\cref{fig:pr-schematic}).

The unwanted AC components are very small compared to the high voltage DC
component. A precise observation of these AC components can be done through a
high-voltage coupling capacitor that removes the DC component from the
high-voltage signal. The remaining AC component can then be looked at without
any attenuation, as opposed to using a frequency-compensated voltage divider
serving both DC and AC stabilization. The absence of attenuation corresponds to
a lower noise in the AC signal path. The AC pick-up capacitor with its
associated protection circuitry is dubbed the ``high-voltage ripple probe'' in
this context. A ``ripple-probe amplifier'' is connected to the high voltage
ripple probe. A high dynamic input impedance of this amplifier provides a cutoff
frequency of the AC pickup system at below \SI{1}{\hertz}. In this way, most of
the regulator feedback loop bandwidth is covered by the low-noise AC path,
resulting in a low total noise of the regulator.

An important design aspect is the required long-term precision of the DC
loop. As outlined in \cref{sec:measChain}, there is already a precision voltage
divider and a precision voltmeter that can provide precision high-voltage
readouts every few seconds. Trying to obtain a similar long-term precision for
the regulator DC stability would have amounted to duplicating the performance of
the very expensive and sophisticated devices above, with very stringent
stability requirements. However, there would be no added value to this effort:
as soon the regulator DC pickup system offers a stability that fulfills the
retarding voltage precision requirements for a few seconds, the high-precision
readouts can be used to apply drift-compensating setpoint corrections of the DC
loop. The DC loop medium-precision divider will be called the "auxiliary
divider" below. Intuitively, one might wonder the feasability of using
precision divider output also for the high voltage post-regulation DC component
acquisition. However, there are several reasons, for keeping these signal paths
separate:
\begin{itemize}
    \item The circuit topology of the auxiliary DC pickup voltage divider input
      can be favorably made different from the conventional voltage divider
      topology. It can be combined with the associated input amplifier in an
      inverting configuration that maps the negative
      \SIrange{0}{-35}{\kilo\volt} HV range into a \SIrange{0}{5}{\volt} range,
      very suitable for analog small signal processing.
    \item The validated precision divider system along with its precision
      multimeter should not be potentially compromised by connecting extra
      amplifiers and other auxiliary components.
    \item The precision divider system has to be calibrated regularly. As long
      as the retarding-voltage regulation has its own divider, it remains
      operative, albeit with reduced long-term stability. The reduced long-term
      stability is sufficient for systematic measurements, for example background
      measurements, that are less sensitive to the retarding potential.
\end{itemize}

\subsection{Shunt regulator principle and vacuum triode shunt device}\label{sec:shuntReg}

A voltage-regulation method known for its fast and precise response is the
so-called shunt regulator method. The load is fed from a roughly regulated
voltage supply via a series resistor. The exact voltage across the load is
adjusted by a controllable dump, or shunt, a device that takes over a part of
the current coming from the primary voltage source across the series
resistor. Such dump devices can be made to act quite fast. They are also very
suitable for primarily capacitive loads, which applies to the KATRIN main
spectrometer. If the shunt device is a controlled current source, the regulator
loop stability will improve with increasing load capacitance \--- a property
that not all voltage regulation principles have. Concerning the high voltage,
shunt regulation was widely used to stabilize the CRT acceleration voltage in
early color TVs in the late 1960s. The dump device in these units was a
high-voltage vacuum shunt triode that would dump the full output power (some
\SI{35}{\watt} at \SI{25}{\kilo\volt}) of the high-voltage generator in case of
a zero beam current (dark picture). Reusing these high-voltage triodes lends
itself well for KATRIN high-voltage control as the tubes can still be purchased
at a low price from the early color TV surplus stocks, and the voltage lies in
the same range of around \SI{25}{\kilo\volt}. Using semiconductors would result
in a complex stacking of multiple devices with the associated fast control
circuitry and the components were much more susceptible to electrical
transients. To assess the usability of these tubes, one must make sure that the
maximum total interference current is lower than the maximum anode current of
this tube type. For this purpose, the vessel was connected to earth ground via
a \SI{1}{\kilo\ohm} resistor and the AC current flowing through that resistor
was measured. This measurement was made with the main electrical installations
of the building and those on the vessel itself, powered through their insulation
transformer, in place and running. The voltage signal across this resistor was
measured using an oscilloscope. The overall peak-to-peak current amplitude was
determined to be within \SIrange{200}{300}{\micro\ampere} with a dominant
\num{50}-\si{\hertz} sinusoidal from the stray fields of the power grid. This is
well below the AC current component that can be compensated when running the
shunt triode at a \SI{0,7}{\milli\ampere} anode current, which will keep the
anode dissipation below \SI{30}{\watt} even at \SI{35}{\kilo\volt}. If a peak
value of \SI{300}{\micro\ampere} is assumed for the interference current, there
will be a \SI{+-400}{\micro\ampere} current control range left for other error
sources, such as an output voltage drift of the primary (commercial)
high-voltage power supply. The remaining control range for voltage deviations is
given by multiplying this current range with the series resistor value. The
series resistor connected between the primary high voltage power supply and the
shunt triode in this system is a \SI{22}{\kilo\ohm} device. The usable voltage
control range of the post-regulation is therefore \SI{+-8.8}{\volt}.

\begin{figure}
  \centering
  \includegraphics[width=0.8\textwidth]{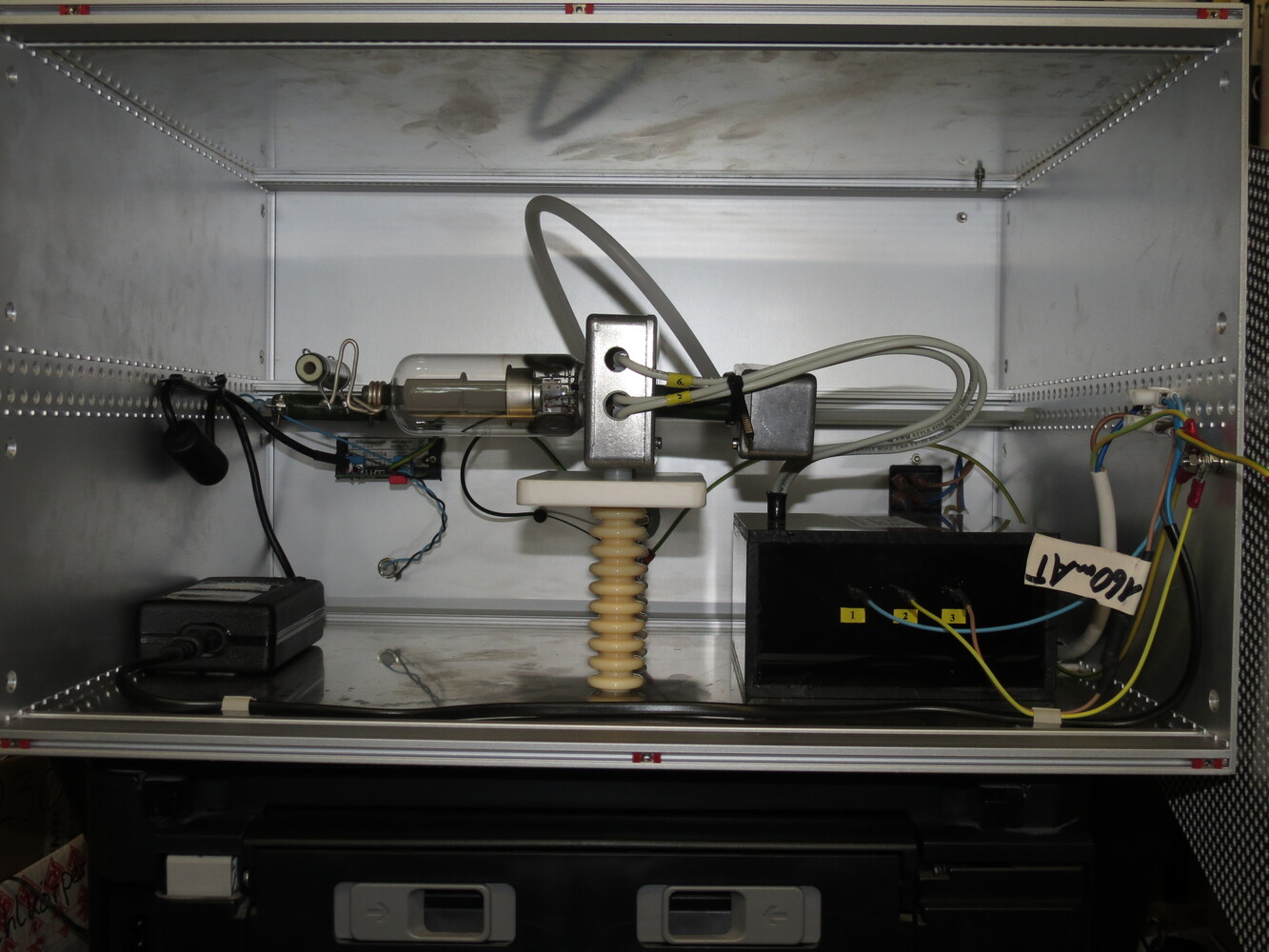}
  \caption{\textbf{View into the 19" rack-mount enclosure of the shunt triode
      unit.} The anode cap of the Philips PD510 triode (on the left) is
    connected to the ground via surge limiting resistors. The cathode and grid
    side along with its fiber-optic-controlled circuitry is floating on the high
    voltage potential (center). Cathode heating and power supply for the control
    circuit is provided by an insulation transformer (on the right)}
  \label{fig:Shunt_unit_interior}
\end{figure}

\begin{figure}
  \centering
  \includegraphics[width=0.8\textwidth]{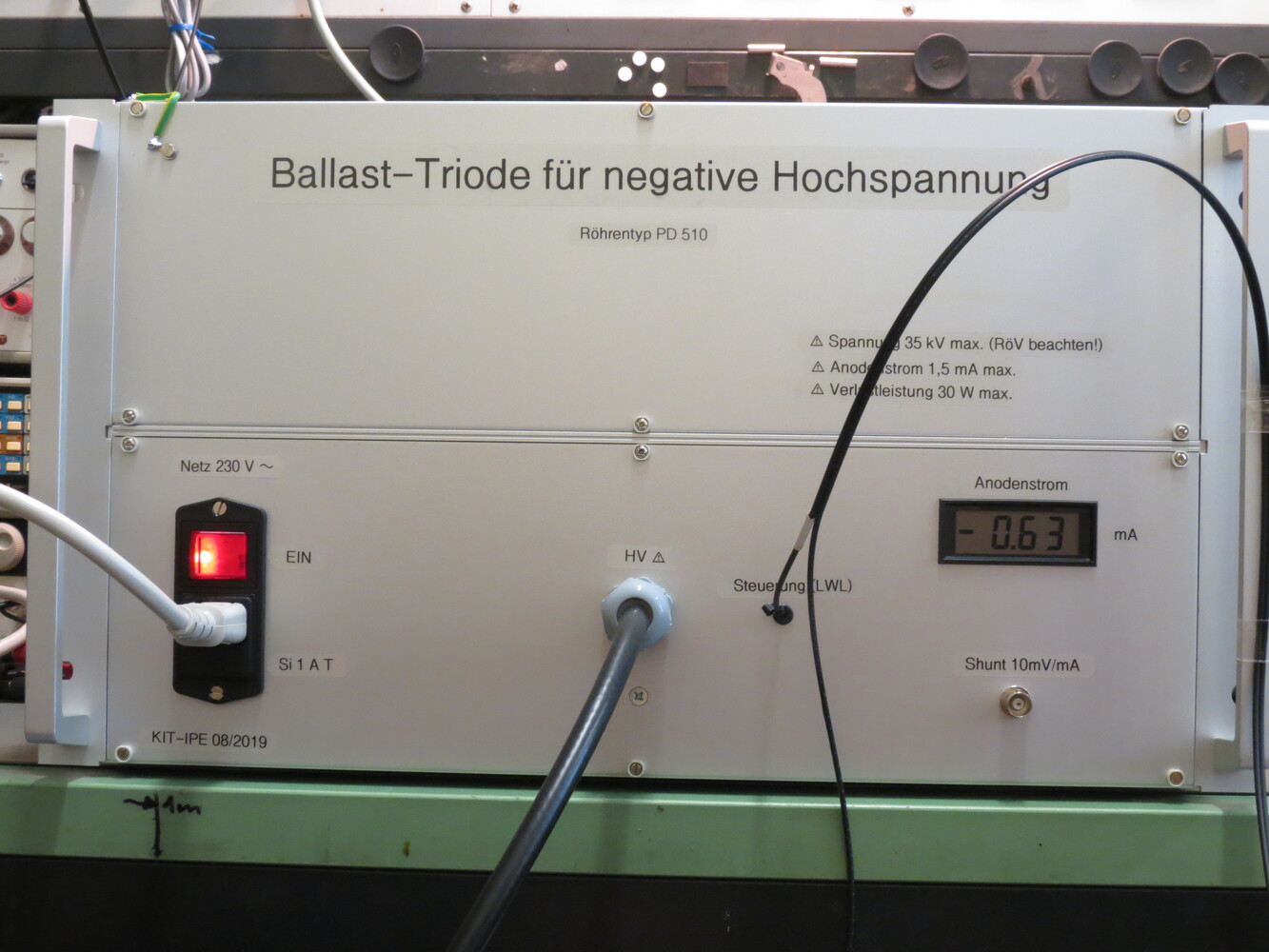}
  \caption{\textbf{Front side of the shunt triode unit.} The anode current can
    be read out visually via the LCD meter and electrically via a BNC receptacle}
  \label{fig:Shunt_unit_front}
\end{figure}

Figures \ref{fig:Shunt_unit_interior} and \ref{fig:Shunt_unit_front} show the
triode shunt unit.

\subsection{Acquisition of the actual high voltage value}

\begin{landscape}
\begin{figure}[t!]
  \centering
  \includegraphics[width=.9\paperwidth]{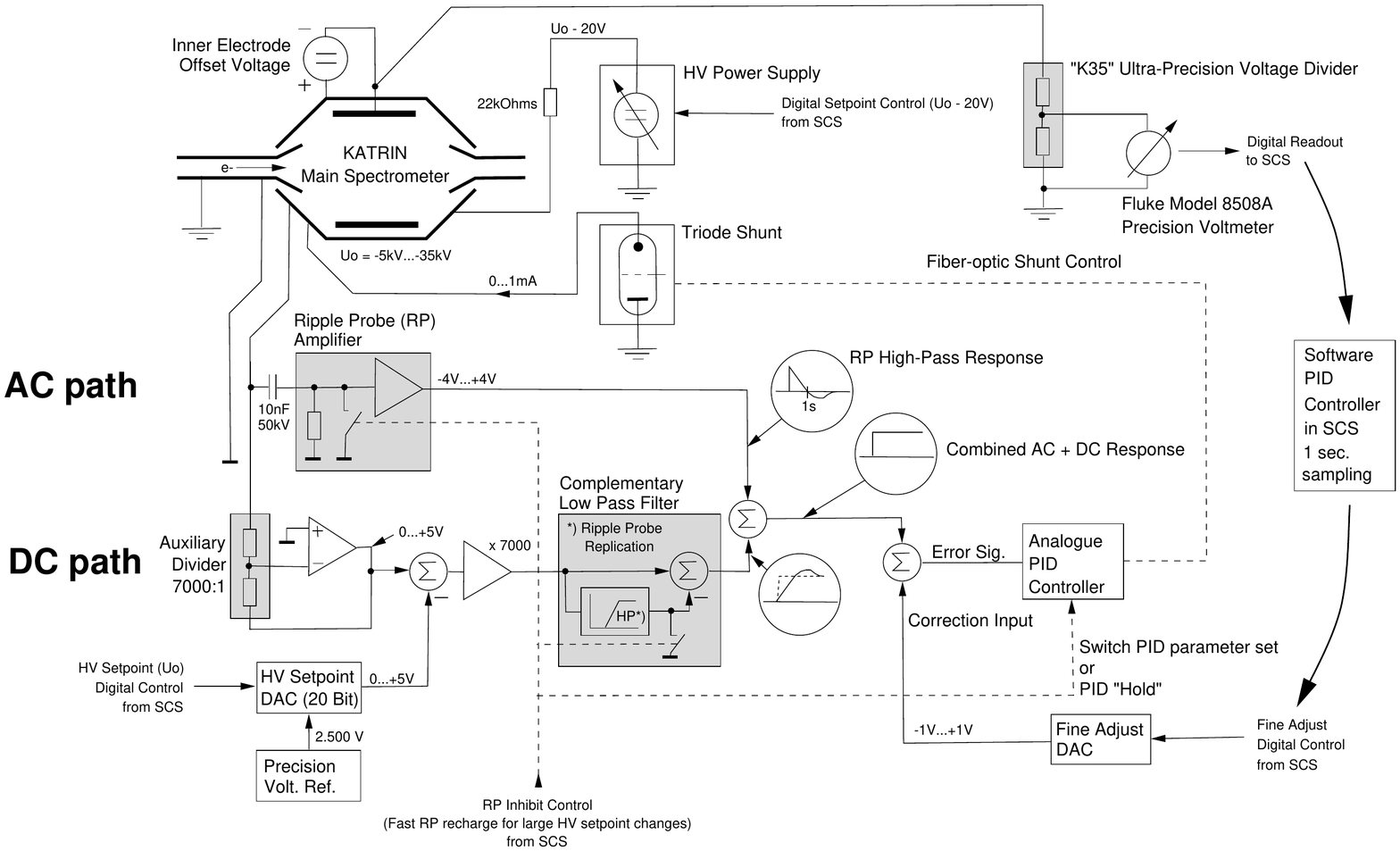}
  \caption{\textbf{Block diagram of the post-regulation system.} A detailed
    block diagram of the post-regulation setup at KATRIN's main spectrometer.}
  \label{fig:blockDiagrammPR}
\end{figure}
\end{landscape}

As outlined in \cref{sec:regulatorStruct}, the acquisition of the actual,
momentary high-voltage value is split into an AC path and a DC path, which will
be described in the following. The high-voltage system in more detail is shown
in \cref{fig:blockDiagrammPR}.

\subsubsection{AC path} \label{sec:rippleProbe}
\begin{figure}[t!]
  \centering
  \includegraphics[width=0.9\textwidth]{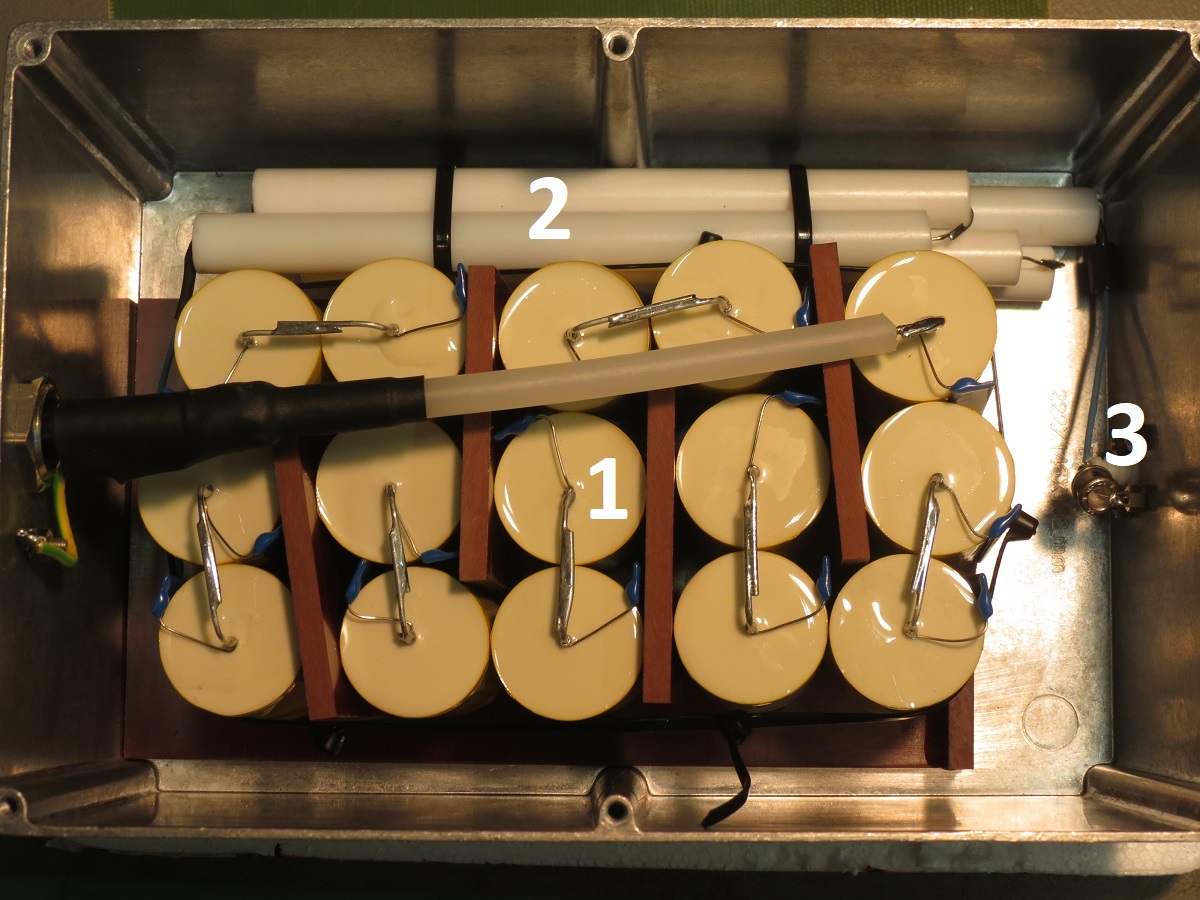}
  \caption{\textbf{Picture showing the interior of the ripple pick-up probe.}
    The device consists of 15 series-connected
    $\SI{150}{\nano\farad}\,/\,\SI{3}{\kilo\volt}$ foil capacitors with
    balancing resistors (1), a surge protection resistor chain (2), and a gas
    discharge surge arrester (3). All components are embedded in silicone
    potting that has not been applied in this photo yet. The width of the
    enclosure is about \SI{21}{cm}.}
  \label{fig:rippleProbe}
\end{figure}

The central element of the AC path is the high-voltage decoupling capacitor that
separates the ripple pickup amplifier from the DC component of the high
voltage. This capacitor must be of very high quality, partial discharges must be
absent at a level of ppm or below. In KATRIN, a series connection of 15
\num{150}-\si{\nano\farad} foil capacitors is used, after experiments of
high-voltage capacitors with paper/oil or ceramic dielectric showed significant
partial discharges. Balancing of the foil capacitor chain is done with 15
resistors of \SI{1}{\giga\ohm} each. Figure \ref{fig:rippleProbe} shows the
interior of the "ripple probe" before the potting with an elastic, transparent
silicone compound.

The ripple pickup amplifier has a very high dynamic input impedance
(\SI{30}{\mega\ohm}), obtained by a bootstrap scheme. This method provides a
high dynamic input impedance while at the same time being able to compensate for
constant DC leakage currents up to several \si{\micro\ampere}.

\begin{figure}[t!]
  \centering
  \includegraphics[width=0.9\textwidth]{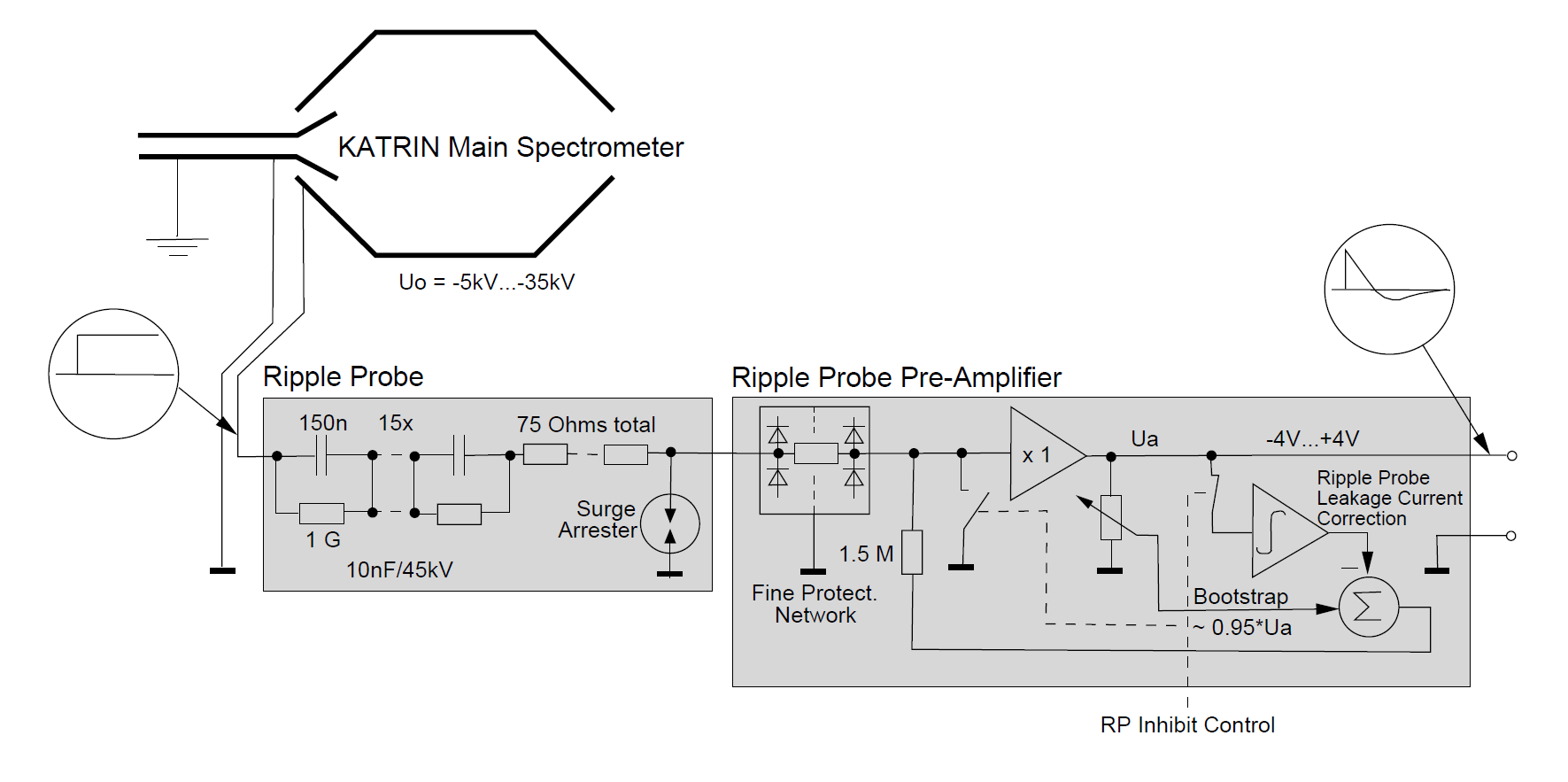}
  \caption{\textbf{Ripple pick-up circuitry.} Block diagram for the ripple probe
  and its pre-amplifier.}
  \label{fig:rippleProbeDiagram}
\end{figure}

With the balanced chain of capacitors, this leakage current is mainly given by
the balancing resistor current.  Another required property of the ripple pickup
amplifier is the ability to withstand high-energy voltage transients at its
input. The pickup capacitor represents a very low impedance for sharp voltage
transients. Such transients can easily occur due to sparking, which may happen
somewhere in the large spectrometer apparatus due to bad insulation vacua and
similar during commissioning A multi-stage protection scheme is employed at the
amplifier input, which consists of a gas surge arrester and three stages of
protection diodes with each stage installed in its shielding chamber. Decoupling
resistors between the stages are chosen to be high enough to exhibit enough
transient damping and low enough not to limit amplifier bandwidth.

The AC path of the post-regulation loop will counteract \textit{any} changes in
the output high voltage, which also applies for \textit{intended} setpoint
steps. This is why an inhibit feature was built into the AC path. If
experimenters want to apply a high voltage setpoint step larger than the AC path
measurement range ($\sim\SI{4}{\volt}_{\mathrm{pp}}$), an inhibit signal can be
asserted that blocks AC smoothing. This function will shorten the ripple
amplifier input to ground, thus allowing the ripple pickup capacitor to quickly
recharge to the new voltage. At the same time, the integrator of the leakage
current compensation will be disconnected, which retains the leakage current
status until the voltage has settled to the new value. However, this method
requires corresponding actions in the remaining regulator loop to maintain
stability. These provisions, such as the adjustment of the PID regulator
parameters, are not yet fully implemented. This is why the experiment has not
yet implemented the AC-path inhibit feature. The ripple amplifier protection
circuitry will still allow for the ripple probe to recharge with an increased
settling time.

The obtained overall settling times and other control-related properties are
given in \cref{sec:dynSetpointCtrl} below.

\begin{figure}[t!]
  \centering
  \includegraphics[width=0.9\textwidth]{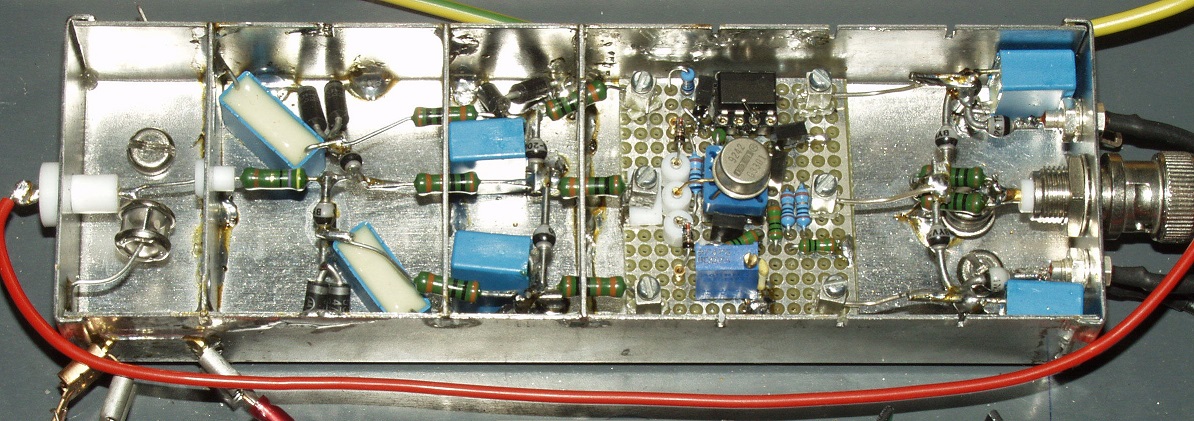}
  \caption{\textbf{Ripple pick-up amplifier with multi-stage transient
      protection.} The HV ripple probe is connected to the terminal on the left
    side. The RF-type enclosure is partitioned into several chambers to
    attenuate transient propagation. The chambers contain (left to right) surge
    arrester, stage 1 protection diodes, stage 2 protection diodes, stage 3
    protection diodes, and the amplifier circuit.}
  \label{fig:pickUpAmplifier}
\end{figure}

\subsubsection{DC path}
\begin{figure}[t!]
  \centering
  \includegraphics[width=0.9\textwidth]{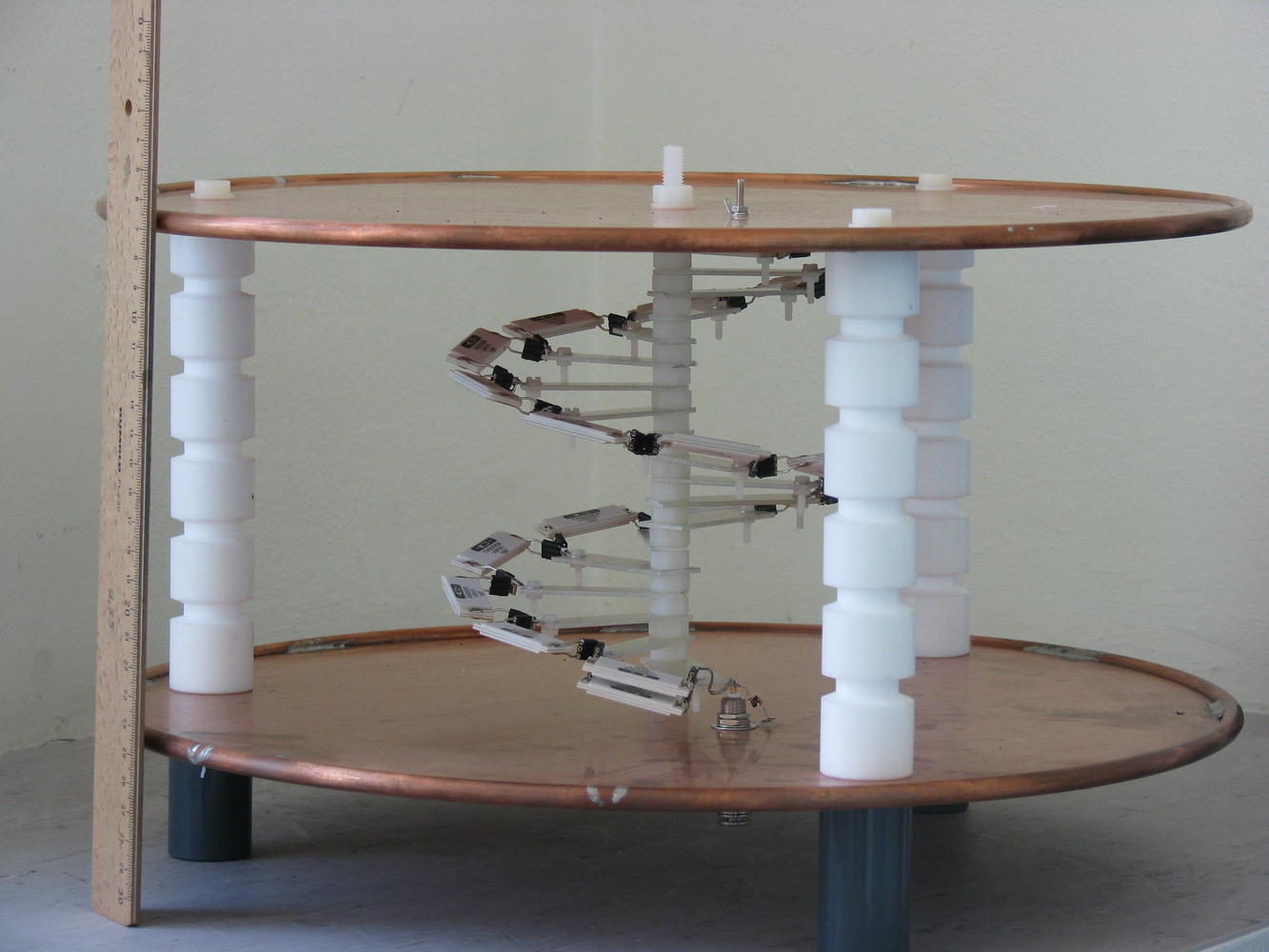}
  \caption{\textbf{Auxiliary voltage divider.} The helical resistor chain of the
    divider is located in between two field-shaping copper plates at
    the top and bottom.}
  \label{fig:auxDivider}
\end{figure}

The DC loop consists of an auxiliary, medium-precision voltage divider, a "zero
offset" operational amplifier stage, a 20-Bit precision setpoint DAC with an
ultra-low noise reference voltage source, and a subtraction stage. The output of
the subtraction stage represents a DC error signal. The error signal is
amplified by two further amplifier stages by a factor equal to the reciprocal of
the voltage-divider ratio (\num{7000}x). Therefore, a \SI{1}{\volt} change at
the output of these stages corresponds to a \SI{1}{\volt} change of the high
voltage, similar to the output of the AC ripple pickup signal path. Ideally, the
sum of the step response signals of the DC and AC pickup paths should form an
ideal step response signal again. This is why the DC path is equipped with a
special low-pass filter formed by a high-pass filter and another subtraction
stage. The high-pass filter consists of a replica of the AC path. Therefore, the
overall DC path step response is complementary to that of the AC path. Figure
\ref{fig:auxDivider} shows the auxiliary divider consisting of a helical
resistor chain and two field-shaping discs. The design objective of this divider
was to be able to hold the required KATRIN retarding voltage accuracy for about
\SI{10}{\second}. This is \--- with some headroom \--- the time the precision HV
readout (cf. \cref{sec:measChain}) needs to produce a valid measurement. The
resistors used for the auxiliary divider are of the Caddock USF200 series. Each
of these resistors consists of two individual resistor elements of opposite
temperature coefficients, glued back to back.

\subsection{Fast-stepping input and DC drift correction}

After merging the outputs of the DC and AC pickup paths into the
\SI[per-mode=symbol]{1}{\volt\per\volt} error signal, it is possible to add a
"fast-stepping" analog control voltage before passing the error signal to the
(PID) regulator. Using this input, it is not only possible to add an intentional
wide-bandwidth waveform to the regulator setpoint, but this input also allows
fine control of the high voltage beyond the resolution of the 20-bit setpoint
DAC.

In the system presented in this paper, this input is used to overlay drift
correction values derived from readouts of the precision voltage divider
readouts as outlined in \cref{sec:regulatorStruct}. A software PI regulator
reads the precision divider output via the precision voltmeter, and creates a
correction input via a special fine-adjustment DAC, by comparing it with the HV
setpoint from the experiment control. We use a \num{10}-bit DAC from the
experiment automation system that provides an output range of
\SIrange{-10}{10}{\volt}. The DAC output voltage is fed to the
post-regulation"fast-stepping" input through a 10:1 voltage divider and can
therefore provide a fine-adjustment range of \SI{+-1}{\volt}.

\subsection{Dynamic setpoint control}\label{sec:dynSetpointCtrl}

For measurements done with the main spectrometer, the retarding potential needs
to be changed synchronously with data taking at the detector. Therefore the
control of the post-regulation is done in close connection with the data
acquisition system for detector data. For most of the measurements (tritium
spectrum spectroscopy, $^{\mathrm{83m}}$Kr conversion electron calibration
measurements), the setpoint of the retarding potential must not only be reached
as fast as possible but also be stable and reproducible on the ppm-level
\cite{KNM1Ana21}.

\begin{figure}
  
  \includegraphics{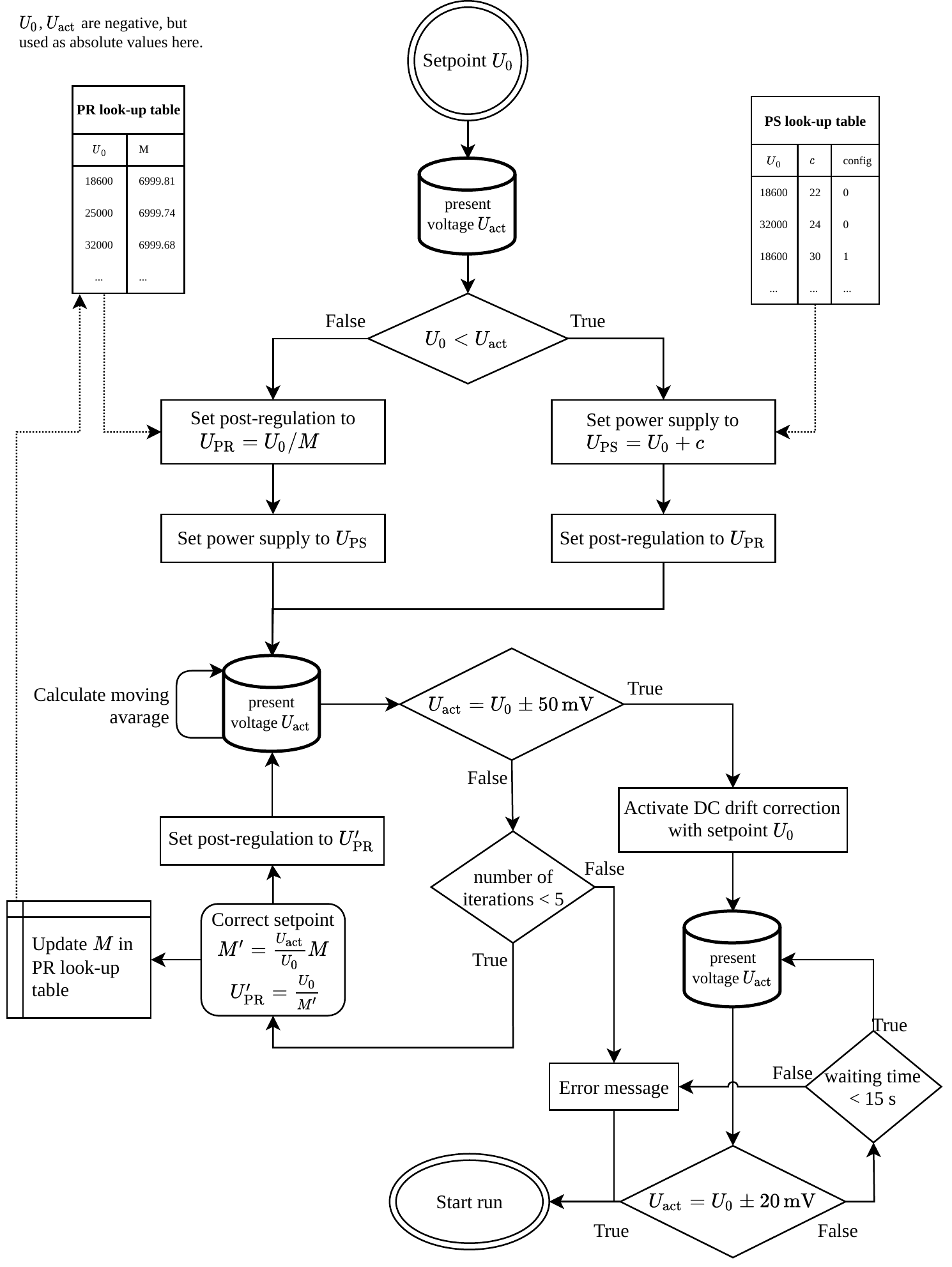}
  \caption{\textbf{Dynamic setpoint control.} Schematic of the logic for a new
    voltage setpoint. ``Start run'' means the start of data taking at the KATRIN
    beamline with a new voltage setpoint.}
  \label{fig:control}
  
\end{figure}

The logic of setting a voltage value is set is shown in detail in
\cref{fig:control}. First, the voltage value is set coarsely
$\mathcal{O}(\SI{~100}{\milli\volt})$. For this operation two inputs are
needed. The setpoint for the primary high voltage supply needs to be more
negative than the required spectrometer voltage to account for the voltage drop
on the shunt regulator series resistor. This offset value $c$ depends on the
individual setup; i.e. how many and which dividers are connected to the main
spectrometer and which power supply is used (as they can have varying absolute
setpoint precision). The offset $c$ needs to be chosen in a way, that the shunt
triode is not overloaded (limited to \SI{30}{\watt}), but still in its optimal
working range of \SIrange{0.5}{0.9}{\milli\ampere}. During the commissioning
measurements, $c$ is determined for each setup, depending on the voltage value
and stored in a look-up table. The second input is the scale factor $M$ of the
auxiliary divider. For new voltage setpoints $U_0$, $M(U_0)$ is determined by
values estimated during commissioning measurements and updated after each
successful setting of the voltage.

After setting the voltage coarsely, the actual voltage is measured with one of
the precision dividers. If necessary, $M$ is adjusted, until the voltage value
has reached its setpoint within $\SI{+-50}{\milli\volt}$. The DC drift
correction is subsequently activated, regulating any remaining deviation and
forcing the voltage to the desired $U_0$. Usually during data taking a loop that
determines and stores the actual $M$ continuously (not drawn in
\cref{fig:control}) is activated. For small voltages steps $< |\SI{50}{\volt}|$,
the most recent value of $M$ is used, instead of the $M(U_0)$ from the look-up
table. For this step-size, changes over time (mainly temperature related) have
higher relevance than the voltage dependency of the auxiliary divider.

Due to the dynamic logic of the setpoint algorithm, the overall settling time
for each voltage setpoint depends on the overall step size, the repetition of
voltage steps, and especially the required precision. For $^{\mathrm{83m}}$Kr
measurements it takes on average \SIrange{22}{25}{\second}, depending on the
step size; the smaller the overall step size, the longer the average settling
time. For standard tritium beta scans, the average step size takes
\SI{29}{\second} on average and is larger than for $^{\mathrm{83m}}$Kr
measurements. One dominant factor of the settling time is the averaging
time over multiple measurements to ensure the setpoint precision before starting
a physics measurement. During these measurements, data is taken continuously at
the detector, but separated into ``usable'' periods with start/stop time
stamps. In the analysis, this time can be recovered again by using the measured
voltage values to redefine the start/stop time of a physics measurement. With
this analysis, the average settling time is reduced for beta scans to
\SI{23}{\second} without losing any precision.

\subsection{X-ray safety}

The shunt triode operating at anode voltages of up to \SI{35}{\kilo\volt} is a
source of stray X-ray radiation and is subject to radiation protection
legislation. Appropriate shielding must be installed and official permission is
required prior to operation. \num{35}-\si{\kilo\electronvolt} X-rays are
relatively soft and are very effectively shielded by the
\num{2.5}-\si{\milli\meter} stainless steel walls of the post-regulation
electronics cabinet. The cabinet door is fitted with a switch that will cut the
power to the shunt triode control and cathode heating when the door is
opened. LED indicators of the post-regulation electronics and a cathode current
meter can be read through lead glass windows installed in the cabinet door.
After a review of the cabinet and radiation measurements by an authorized
radiation safety engineer, a permit was issued by the authority,
Regierungspräsidium Karlsruhe.

\section{Performance measurements}

During the development of the post-regulation system, several quality and
performance checks were carried out \cite{PhDKraus2016,PhDRest2019}. At present
four different methods allow for a measurement of the remaining ripple on the
main spectrometer retarding voltage on different time scales. In this section,
the performance of the final implementation, for the neutrino mass measurements,
is evaluated with these methods.

\subsection{Ripple pick-up probe}\label{sec:rippleProbeMeasurement}

\begin{figure}[t]
    \centering
    \includegraphics[width=\textwidth]{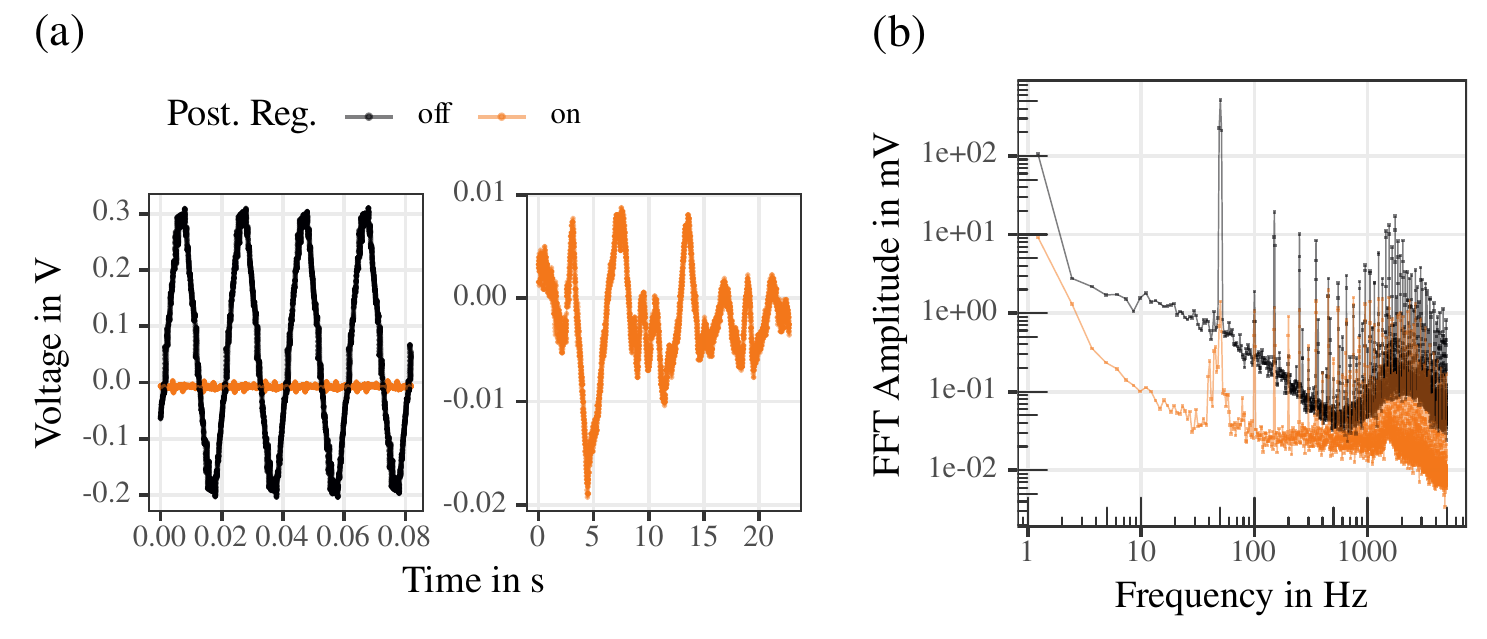}
    \caption{\textbf{Oscilloscope readout of ripple pick-up probe.} Two traces
      taken with the ripple pick-up probe at \SI{-18.3}{\kilo\volt} main
      spectrometer potential are shown. In (a) a measurement without the
      post-regulation at a sample rate of \SI{10}{\kilo\hertz} and with active
      post-regulation at a sample rate of \SI{250}{\hertz}, is shown (please
      note the different scales). Plot (b) shows the Fourier spectrum for both
      cases.}
    \label{fig:rippleProbeTrace}
\end{figure}

The ripple pick-up probe (cf. \cref{sec:rippleProbe}) allows for a measurement
of the AC noise on the retarding potential. Traces of the ripple pick-up probe
at \SI{-18.3}{\kilo\volt} are shown in \cref{fig:rippleProbeTrace}. Without the
post-regulation, the AC noise is dominated by a sinusoidal \SI{50}{\hertz}
signal, with amplitudes of about \SI{0.25}{\volt}. With the post-regulation
active the sinusoidal noise is suppressed and only fluctuations within less than
\SI{0.03}{\volt} remain.

To analyze the AC components further, \num{100} traces at a sampling rate of
\SI{10}{\kilo\hertz} are taken. Each trace was limited to \num{8192} samples. To
remove any remaining DC part, the mean voltage value is first calculated for
each trace, which is then subtracted from each voltage value within one trace. A
fast Fourier transformation with a Hamming window (window function) is applied
on the net trace and plotted in the right plot of
\cref{fig:rippleProbeTrace}. Comparing the Fourier spectrum with and without
active post-regulation, one can see a noise reduction on a wide range of
frequencies. The largest component at \SI{50}{\hertz} is suppressed by two
orders of magnitude.

\subsection{Conversion electron lines from $^{\mathrm{83m}}$Kr}

Measurements with conversion electron lines from $^{\mathrm{83m}}$Kr provide an
independent method to the ripple pick-up probe measurements. The conversion
electrons from $^{\mathrm{83m}}$Kr are equally sensitive to the ripple as the
beta decay electrons, whereas the ripple pick-up probe is only attached to the
outside of the main spectrometer vessel. Therefore the electrons are an ideal
tool for an independent cross-check of the post-regulation performance.

Measurements of $^{\mathrm{83m}}$Kr conversion electron lines are a standard
method at KATRIN \cite{kryptonHvCal2017, Venos2018, Altenmueller2020}. The
measurements discussed here were performed with the \textit{condensed Krypton
source} (CKrS) during a measurement campaign described in
\cite{PhDFulst2020}. They consist of measurements from three conversion electron
lines, following the \num{32}-\si{\kilo\electronvolt} gamma transition, but from
different shells: K-32 at \SI{17.8}{\kilo\electronvolt}, L$_3$-32 at
\SI{30.5}{\kilo\electronvolt} and the line doublet N$_2$N$_3$-32 at
\SI{32.1}{\kilo\electronvolt}.

For a typical transmission function scan of a conversion electron line with the
main spectrometer, the retarding potential is changed in small steps
(\SIrange{0.1}{0.5}{\volt}) around the expected line position. The width of the
measured transmission function depends on the spectrometer resolution and the
natural line width. Each transmission function has a retarding potential at
which the rate is half of the rate compared to full transmission, called the
middle of transmission. The rate in the transmission region is correlated with
the retarding potential. For small variations of the retarding potential ($<
\SI{1}{\volt}$, depending on the width of the transmission function) around the
middle of transmission, the measured electron rate at the detector is linearly
dependent on the retarding potential.

By measuring the shape of the transmission function in a dedicated measurement
the linear dependency of the rate on the retarding potential can be
estimated. In a second step, the rate at the middle of transmission is
measured. Assuming a perfectly stable source any fluctuations of the rate can
then be accounted as fluctuations of the retarding potential.

The measurements focus on the largest noise part, coming from the
\SI{50}{\hertz} power grid interference. The dedicated grid synchronization
signal at KATRIN's focal plane detector (FPD) measures the
$\approx\SI{50}{\hertz}$ frequency of the mains power and outputs a
synchronization pulse (called the \textit{grid sync pulse}) for each new mains
power period (cf. p. 79, \cite{DesignReport21}). With this, each event measured
at the FPD has a relative time to the start of a new mains power period
$t_{\mathrm{rel}}$.

\begin{figure}[t!]
    \centering
    \includegraphics[width=0.49\textwidth]{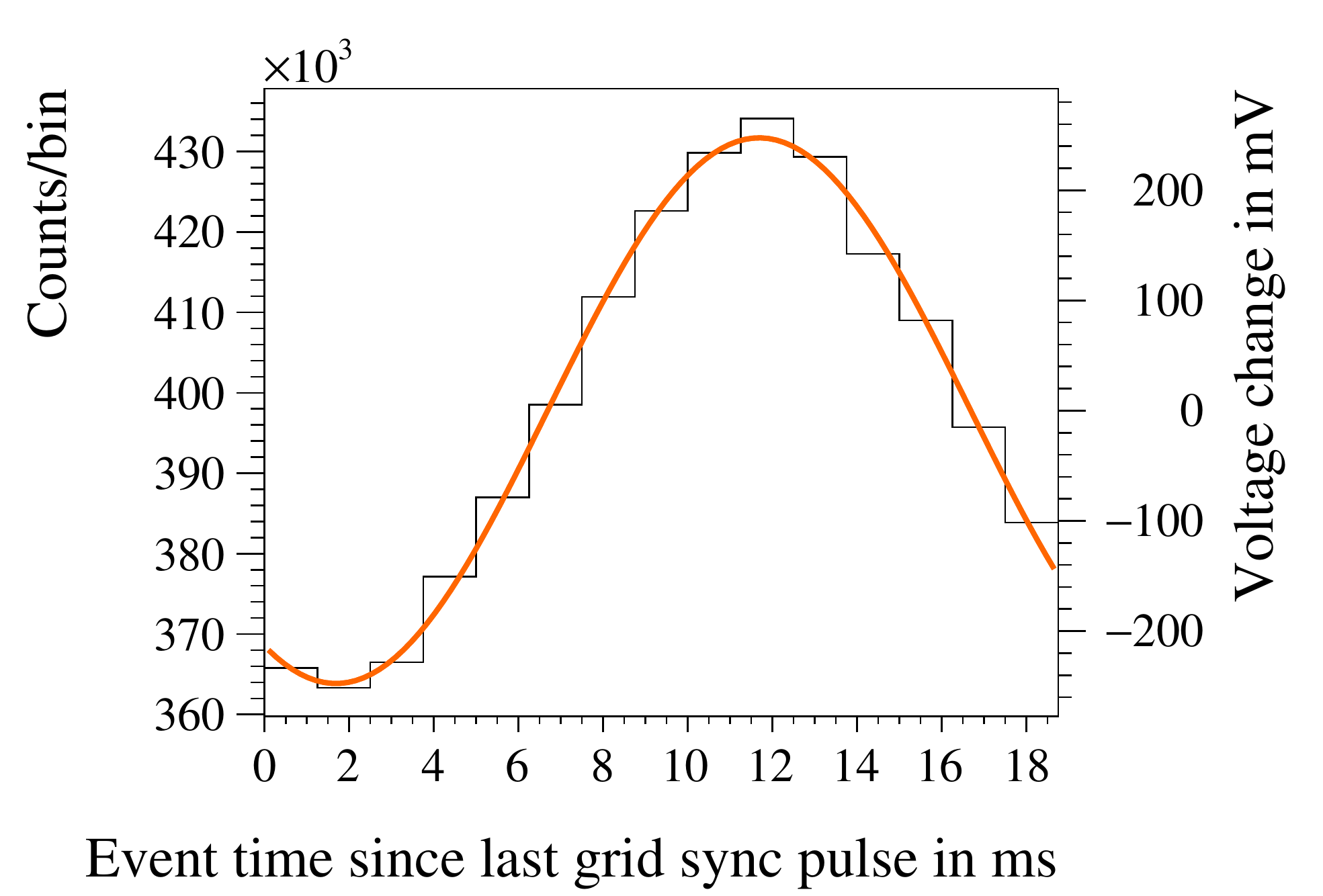}
    \hfill
    \includegraphics[width=0.49\textwidth]{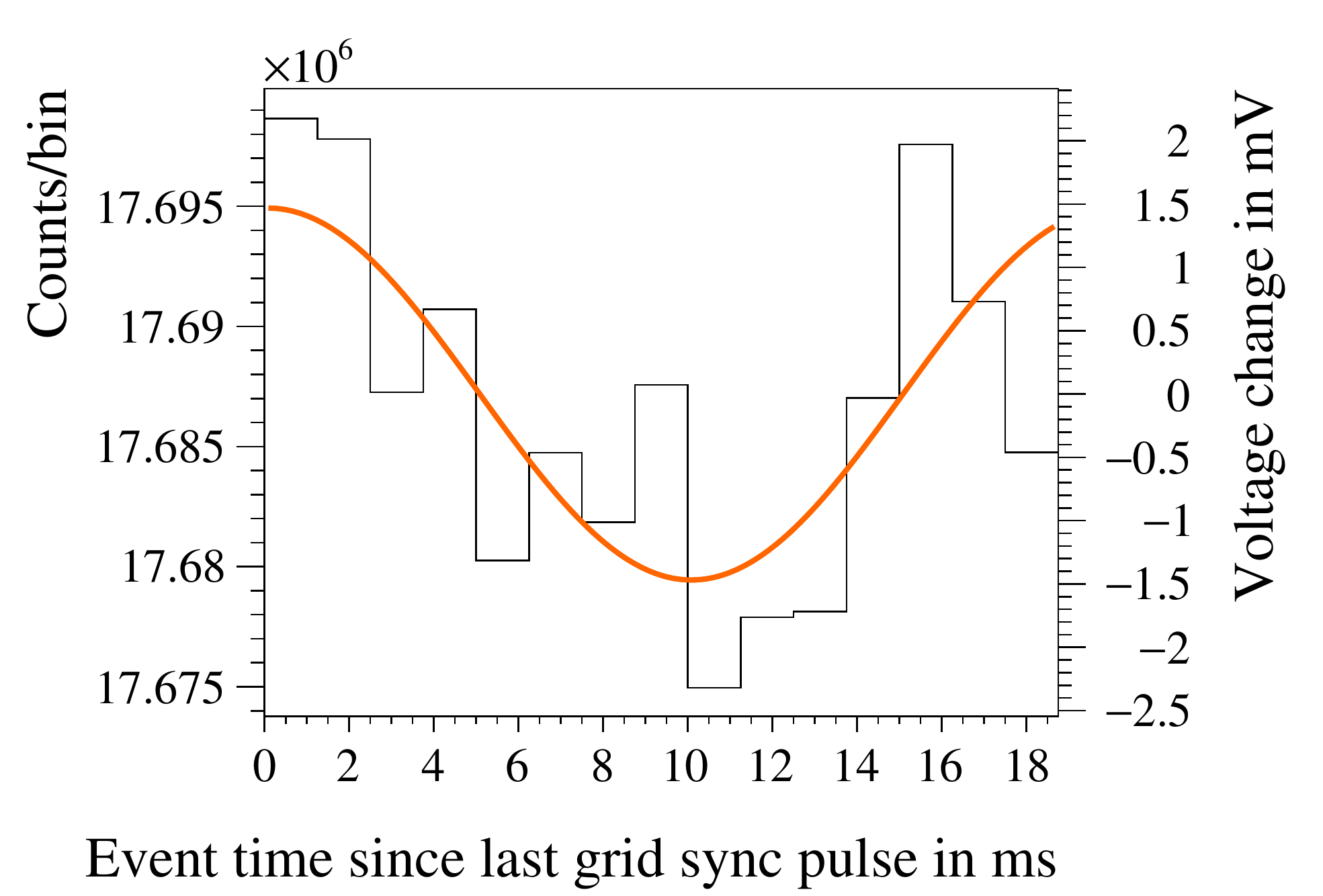}
    \caption{\textbf{Line-center measurements.} Results of the line-center
      measurement with the L$_3$-32 line. The left plot shows a
      \num{10.5}-minute measurement without post-regulation. A clear sinusoidal
      ripple with an amplitude of \num{33922+-231} counts, transforming to a
      voltage ripple of \SI{247+-2}{\milli\volt} is visible. The
      \num{7.5}-\si{hour} measurement with active post-regulation is shown in
      the right plot. Here the sinusoidal structure is not as
      significant. Still, a fit results in an amplitude of \num{7741+-1592}
      counts transformed to a voltage ripple of \SI{1.5+-0.3}{\milli\volt}.}
    \label{fig:ripple50Hz}
\end{figure}

Measuring the rate at the middle of transmission and stacking the measured
events at the detector along $t_{\mathrm{rel}}$ modulo \SI{20}{\milli\second}
reveals directly any \SI{50}{\hertz} noise seen by the electrons on their way
through the spectrometer. In the following, this type of measurement is called
the \textit{line-center} method. The result of such a measurement is plotted in
\cref{fig:ripple50Hz}. The left plot is without post-regulation. Here the rate
at the middle of transmission is following a \SI{50}{\hertz} sinusoidal
ripple. In a dedicated transmission function scan, the slope around the middle
of transmission is determined. With the slope, the rate change can be
transformed to a voltage ripple of about \SI{0.25}{\volt}. In the right plot,
with active post-regulation, the sinusoidal structure is not as distinct,
despite a 43-fold increased measurement time. This clearly shows the ability of
the post-regulation to decrease the \SI{50}{\hertz} sinusoidal ripple by at
least two orders of magnitude.

By repeating the line-center method at different line positions, a possible
dependency on the size of the retarding potential can be investigated. It also
allows the comparison of the measured ripple amplitude at the ripple probe to
the amplitude measured by the line-center method. To investigate the sensitivity
of the method, measurements with different power supplies and without
post-regulation were performed.

\begin{table}
\centering
\caption{\textbf{Ripple amplitudes without post-regulation.} The amplitudes are
  estimated by two different measurement methods (line-center and ripple probe),
  for different power supplies. The ripple amplitude as measured with
  $^{\mathrm{83m}}$Kr conversion electrons at the detector (line-center) and of
  the ripple probe is determined by fitting a sinusoidal with free amplitude.}
\begin{tabular}{lrcrr}
\toprule
Line & Ret. pot (kV) & Power supply & Ripple probe (mV) & Line-center (mV)\\
\midrule
&& A & \num{225\pm1} & \num{219\pm4}\\
K-32 & \num{-17.8} & B & \num{234\pm1} & \num{222\pm4}\\
&& C & \num{210\pm1} & \num{204\pm4}\\
\midrule
&&A & \num{248\pm1} & \num{246\pm2}\\
L$_{\mathrm{3}}$-32& \num{-30.5}& B & \num{258\pm1} & \num{248\pm2})\\
&& C & \num{229\pm1} & \num{231\pm2}\\
\midrule
N$_2$N$_3$&\num{-32.1}& C & \num{233\pm1} & \num{237\pm6}\\
\bottomrule
\end{tabular}
\label{tab:RippleComp}
\end{table}

Without post-regulation the ripple amplitudes as measured with the ripple probe
and the line-center method are in good agreement, see results in
\cref{tab:RippleComp}. The only exception is the measurement with power supply
B. Here the deviation between both methods is larger than for devices A and C
but most likely due to a drift of the device between the ripple probe and the
line-center measurement. This device was added to the measurements to have a
larger variety of devices but is not used in normal measurement campaigns, due
to its increased instabilities and fluctuations compared to devices of the same
type. Note that any drift of the power supply would be compensated by the
post-regulation within its regulation limit (c.f. \cref{sec:shuntReg}).

The variety of tests show that the ripple probe measurement and the line-center
method yield very similar results for the amplitude of the \SI{50}{\hertz}
noise. This proves, at least for the \SI{50}{\hertz} noise, that the ripple
probe readout matches the ripple seen by the electrons inside the main
spectrometer.

With post-regulation, the comparison of both methods is more difficult, since
here the sinusoidal \SI{50}{\hertz} noise is not the dominant part
(c.f. \cref{fig:rippleProbeTrace}) and the line-center method is only sensitive
to the \SI{50}{\hertz}. Still, both methods show the clear mitigation of the
\SI{50}{\hertz} ripple by the post-regulation.

\subsection{Precision high voltage divider}

\begin{figure}
  \includegraphics{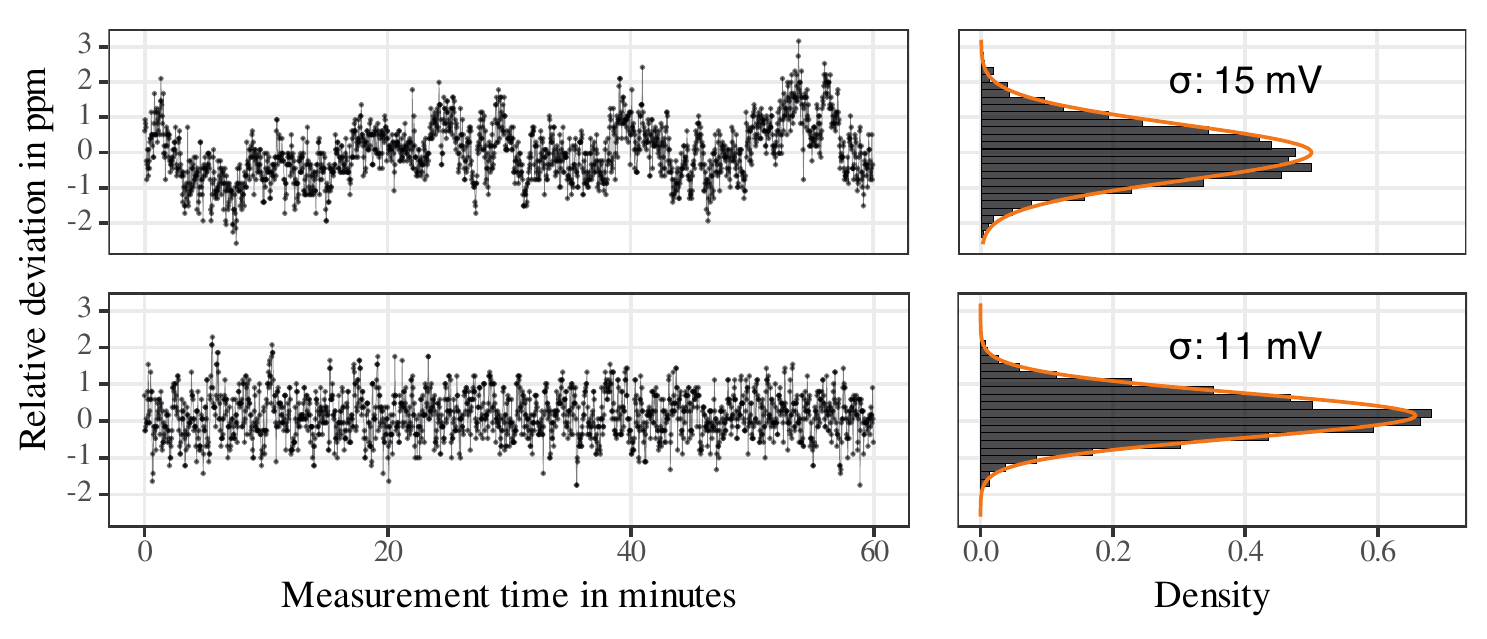}
  \caption{\textbf{Stability measurements.} Measurement with the precision high
    voltage divider K35 over \SI{1}{\hour} at a retarding potential of
    \SI{-18.6}{\kilo\volt}. The upper plot shows the performance without the DC
    drift correction, the lower with the DC precision loop active.}
  \label{fig:K35longterm}
\end{figure}

As described in \cref{sec:measChain}, a continuous measurement of the retarding
potential with a readout rate of \SI{0.5}{\hertz} of a precision high voltage
divider is in place. The stability of an exemplary one-hour measurement at a
fixed retarding potential of \SI{-18.6}{\kilo\volt} is shown in
\cref{fig:K35longterm}. Without the DC drift correction a random-walk-like
structure within \SI{+-3}{\ppm} is visible. With active DC drift correction, the
measured voltage has a standard deviation of \SI{11}{\milli\volt}. This
Gaussian-like distribution is valid at all times from minutes to two weeks. It
is only limited by the long-term stability of the divider and voltmeter.

To cover the gap in frequency between the sensitivity range of the high voltage
divider and the ripple probe measurement, an additional measurement system, the
``fast-measurement system'', was temporarily installed \cite{PhDRest2019}.

The fast-measurement system measures the vessel potential as a difference
voltage to a reference potential. The reference potential is supplied by a
sufficiently stable power supply\footnote{The power supply used for this setup
is a FuG HCP 70M-35000. Its stability is verified to be \SI{2}{ppm} over
\SI{8}{\hour} \cite{PhDRest2019}.}. For instance, when aiming for a measurement
in the range of \SI{-18600}{\volt}, the reference potential is set to a value
close to this voltage (e.g. \SI{-18590}{\volt}), such that the difference
voltage is in the order of \SI{10}{\volt}. This voltage range can be measured
directly with a voltmeter, without a high-voltage divider inbetween. Therefore a
\num{6.5}-digit measurement offers a more than sufficent precision
(\si{\milli\volt} resolution is sufficient) and allowing a fast and precise
measurement, with a sample rate of \SI{3}{\hertz}\footnote{The voltmeter used
for this measurement is a Fluke 8846A.}. The reference potential is monitored
with a high voltage divider, scaling the reference potential down by a factor of
about \num{2000} and measuring it with \num{8.5}-digit precision at a readout
rate of \SI{1/4}{\hertz}. In the analysis, the vessel potential is then
estimated by adding the difference voltage to an interpolated value of the
reference potential.

\begin{figure}
  \includegraphics{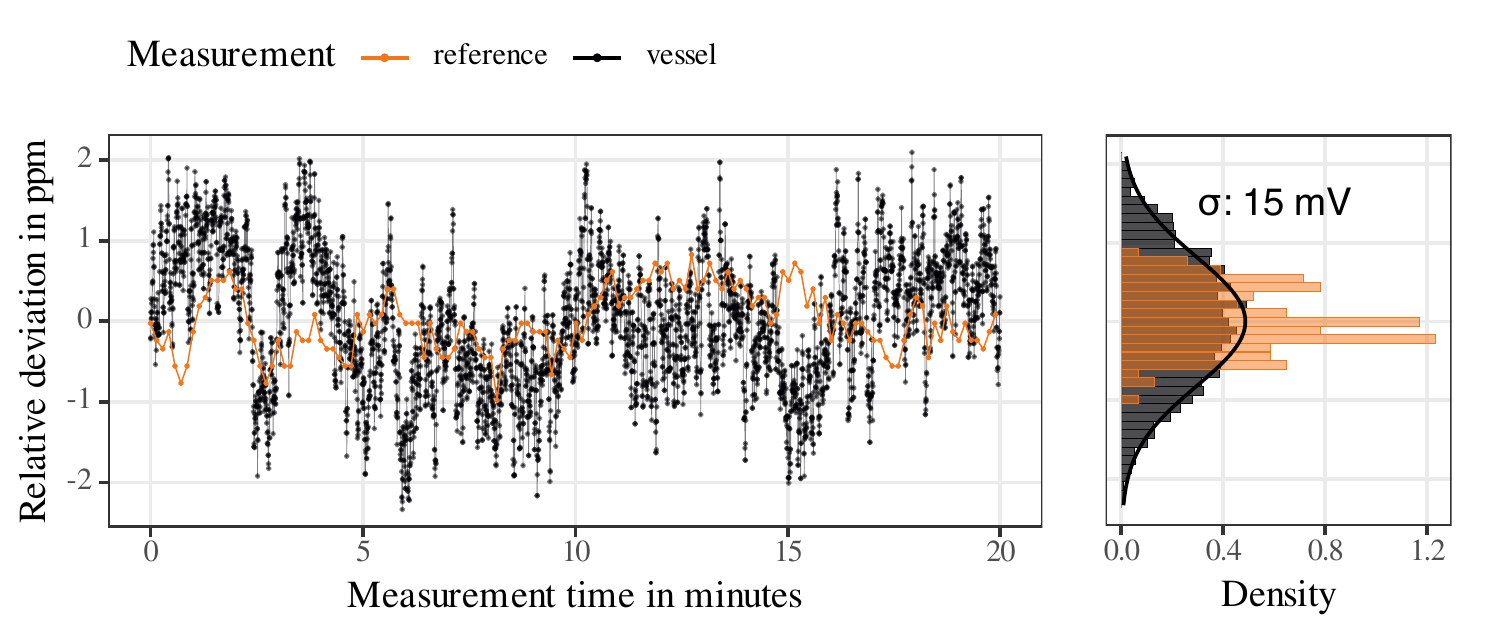}
  \caption{\textbf{Fast voltage measurement.} The relative deviation of the
    reference potential as measured by the high voltage divider K35 is plotted
    in orange. The relative deviation of the vessel potential (at
    \SI{-18.6}{\kilo\volt}) is plotted in black. The vessel potential is the sum
    of reference potential and difference potential.}
  \label{fig:fastMeas}
\end{figure}

One exemplary measurement over \SI{20}{\minute} is shown in
\cref{fig:fastMeas}. During the measurement, the DC drift correction was not
used. The relative deviation over time is similar to the one measured with the
standard measurement setup (upper plot in \cref{fig:K35longterm}); the voltage
is changing randomly within \SI{+-3}{\ppm}.

\subsection{Evaluation over the full frequency range}

As presented above with all three methods: high-voltage divider,
fast-measurement system and ripple probe, the stability of the retarding
potential was investigated at a fixed value of \SI{-18.6}{\kilo\volt}. To
evaluate the stability on different time scales even further, the Allan variance
method \cite{allan66} is used.

The Allan variance provides a tool to estimate the stability of a sensor due to
noise sources on all time scales, limited by the measurement frequency and the
sample size. It is calculated as:
\begin{align}
  \sigma_y^2(\tau) = \frac{1}{2} E\left((\bar{y}_{n+1} - \bar{y}_n)^2\right).
\end{align}
The data $y(t)$ is divided into $n$ parts with $n$ being a multiple of the
measurement frequency. $\tau$ is the time length of each part. For all parts the
mean value, and its deviation to the subsequent mean value, is determined. For
each $\tau$ the expectation value $E$ of the deviations is
determined. Calculating the square root of the Allan variance gives the Allan
deviation.

\begin{figure}
  \includegraphics{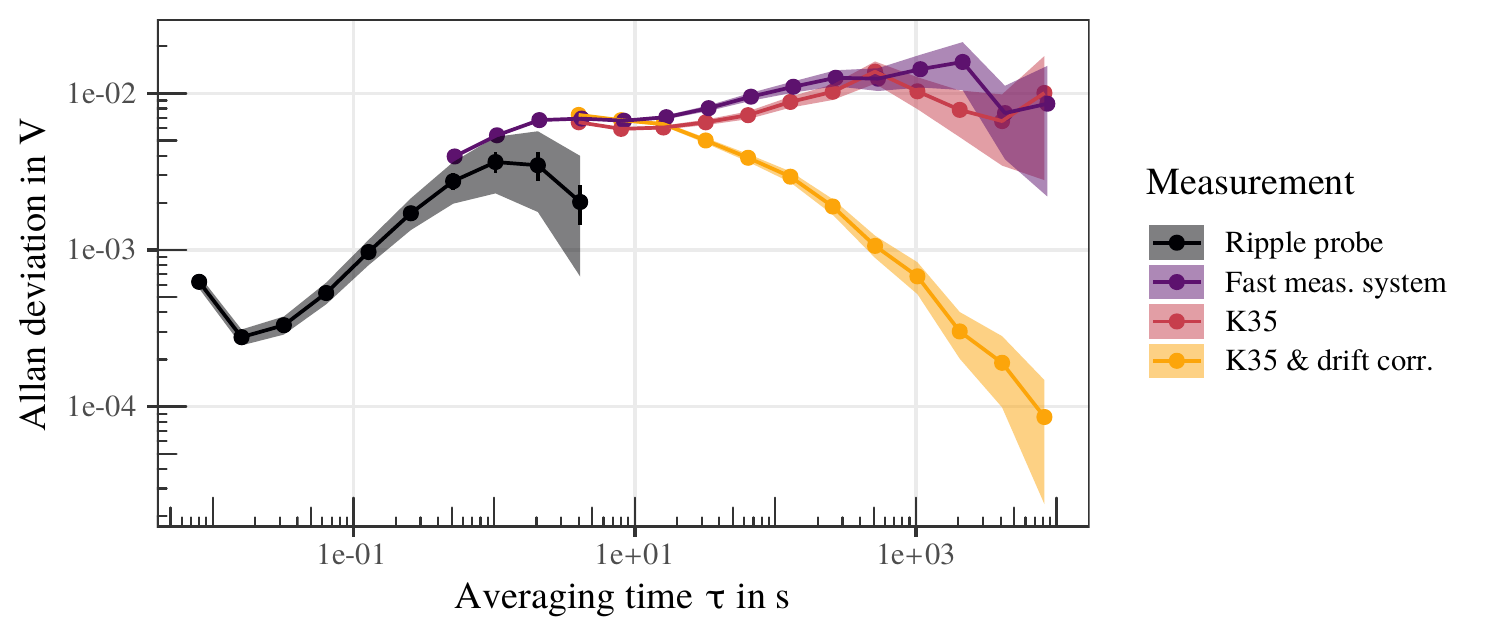}
  \caption{\textbf{Allan deviation.} At a retarding potential of
    \SI{-18.6}{\kilo\volt} the Allan deviation was determined for different
    measurement types. In all four cases, the post-regulation system was active
    without the DC drift correction. Except for the data shown in yellow, here
    the DC drift correction was active. The readout rate for the K35 voltmeter
    is \SI{1/2}{\hertz} and for the fast-measurement system \SI{3}{\hertz}. For
    all three the sample interval spans over \SI{11}{\hour}. For the ripple
    probe, the avarage Allan deviation over \num{1004} traces is plotted. Each
    trace is taken at a sample rate of \SI{250}{\hertz} over \SI{33}{\second},
    the ripple probe is not limited to \SI{250}{\hertz}
    (c.f. \ref{sec:rippleProbeMeasurement}).}
  \label{fig:allan}
\end{figure}

For all three measurement methods, the Allan deviation is determined and plotted
in \cref{fig:allan}. Comparing the measurement with the fast-measurement system
and the ripple probe measurement one can see that the sensitivity of the
fast-measurement system starts where the ripple probe's sensitivity fades. It is
important to note that the instability increases for longer time scales, so the
part measured by the ripple probe but not seen by the fast-measurement system is
more stable. This is again true for frequencies not measured by the high-voltage
divider K35, here the fast-measurement system shows an overall higher
stability. The fast instabilities measured by the fast-measurement system, are
far below our stability requirement. It is very well justified to
have only the K35 measurement in place for continuous monitoring of the high
voltage during neutrino mass measurements at KATRIN.

The measurements with the fast-measurement system and with the high-voltage
divider K35 were both performed without the DC drift correction. The yellow
curve in \cref{fig:allan} shows the performance with the DC drift
correction. The DC drift correction removes the drifts on longer time
scales ($>$ minutes), as already visible in \cref{fig:K35longterm}.

\section{Conclusion}

A custom-designed high-voltage system for KATRIN's main spectrometer has been
developed, implemented, and tested. It outperforms the accuracy requirements set
in the design report \cite{DesignReport05} by providing a retarding high voltage
with sub-ppm precision and ppm-trueness at timescales from microseconds up to
several weeks.

This solution comprises a nested regulation structure starting with a
non-attenuating ripple pickup for low-noise smoothing of the retarding voltage,
employing a medium-accuracy auxiliary voltage divider for intermediate
timescales (seconds) and a drift compensation. The regulation structure deals
with significant interference sources imposed by the large spectrometer vessel,
which is not only openly exposed to external electromagnetic influence but also
hosts its own isolated AC power net. Fast fine control of the high voltage is
obtained by a ballast triode shunt regulator seen before in 1960s color TV
sets. For the drift compensation, precision high-voltage dividers are in place,
allowing one to trace the retarding voltage back to absolute metrological standards.
The performance of the system was proven with independent methods, including
$^{\mathrm{83m}}$Kr conversion electron measurements.

During ongoing neutrino mass measurement campaigns \cite{KNM12019,KNM22021} the
high-voltage regulation system is continuously running and providing a retarding
potential with ppm-accuracy. Up to now, and also to be expected for future
campaigns, the systematic uncertainty of the retarding potential is negligible
for the neutrino mass measurement of KATRIN.

\acknowledgments

The authors would like to express their gratitude to the members of the KATRIN
collaboration, in particular G. Franklin and L. Thorne for the GridSynch box,
and the slow control and detector run control task group, with special thanks to
F. Fränkle and T. Höhn. We acknowledge the support of Helmholtz Association
(HGF), Ministry for Education and Research BMBF (05A20PMA), Deutsche
Forschungsgemeinschaft DFG (Research Training Group GRK 2149) and the Department
of Energy through grants DE-FG02-97ER41020, DE-FG02-94ER40818, DE-SC0004036,
DE-FG02-97ER41033, DE-FG02-97ER41041, {DE-SC0011091 and DE-SC0019304 and the Federal Prime Agreement DE-AC02-05CH11231} in the United States.

\bibliographystyle{JHEP}
\bibliography{references.bib}

\providecommand{\href}[2]{#2}\begingroup\raggedright\begin{thebibliography}{10}

\bibitem{DesignReport05}
{KATRIN collaboration}, \emph{{KATRIN} design report},  {FZKA} scientific
  report \href{http://bibliothek.fzk.de/zb/berichte/FZKA7090.pdf}{7090},
  Forschungszentrum Jülich (2005),
  \href{https://doi.org/10.5445/IR/270060419}{DOI}.

\bibitem{DesignReport21}
{\scshape KATRIN} collaboration, \emph{The design, construction, and
  commissioning of the katrin experiment},
  \href{https://doi.org/10.1088/1748-0221/16/08/t08015}{\emph{Journal of
  Instrumentation} {\bfseries 16} (2021) T08015}.

\bibitem{Marsteller2020}
A.~Marsteller, B.~Bornschein, L.~Bornschein, G.~Drexlin, F.~Friedel, R.~Gehring
  et~al., \emph{Neutral tritium gas reduction in the katrin differential
  pumping sections},
  \href{https://doi.org/https://doi.org/10.1016/j.vacuum.2020.109979}{\emph{Vacuum}
  {\bfseries 184} (2021) 109979}.

\bibitem{Beamson1980}
G.~Beamson, H.Q.~Porter and D.W.~Turner, \emph{The collimating and magnifying
  properties of a superconducting field photoelectron spectrometer},
  \href{https://doi.org/10.1088/0022-3735/13/1/018}{\emph{Journal of Physics E:
  Scientific Instruments} {\bfseries 13} (1980) 64}.

\bibitem{Lobashev1985}
V.M.~Lobashev and P.E.~Spivak, \emph{A method for measuring the electron
  antineutrino rest mass},
  \href{https://doi.org/10.1016/0168-9002(85)90640-0}{\emph{Nuclear Instruments
  and Methods in Physics Research Section A: Accelerators, Spectrometers,
  Detectors and Associated Equipment} {\bfseries 240} (1985) 305}.

\bibitem{Picard1992}
A.~Picard, H.~Backe, H.~Barth, J.~Bonn, B.~Degen, T.~Edling et~al., \emph{A
  solenoid retarding spectrometer with high resolution and transmission for
  {keV} electrons},
  \href{https://doi.org/10.1016/0168-583X(92)95119-C}{\emph{Nuclear Instruments
  and Methods in Physics Research Section B: Beam Interactions with Materials
  and Atoms} {\bfseries 63} (1992) 345}.

\bibitem{amsbaugh2015}
J.~Amsbaugh, J.~Barrett, A.~Beglarian, T.~Bergmann, H.~Bichsel, L.~Bodine
  et~al., \emph{Focal-plane detector system for the katrin experiment},
  \href{https://doi.org/https://doi.org/10.1016/j.nima.2014.12.116}{\emph{Nuclear
  Instruments and Methods in Physics Research Section A: Accelerators,
  Spectrometers, Detectors and Associated Equipment} {\bfseries 778} (2015)
  40}.

\bibitem{K3509}
T.~Th{\"u}mmler, R.~Marx and C.~Weinheimer, \emph{Precision high voltage
  divider for the {KATRIN} experiment},
  \href{https://doi.org/10.1088/1367-2630/11/10/103007}{\emph{New Journal of
  Physics} {\bfseries 11} (2009) 103007}.

\bibitem{K6513}
S.~Bauer, R.~Berendes, F.~Hochschulz, H.~Ortjohann, S.~Rosendahl,
  T.~Th{\"u}mmler et~al., \emph{Next generation {KATRIN} high precision voltage
  divider for voltages up to {65kV}},
  \href{https://doi.org/10.1088/1748-0221/8/10/P10026}{\emph{Journal of
  Instrumentation} {\bfseries 8} (2013) P10026}.

\bibitem{KNM22021}
{\scshape KATRIN} collaboration, M.~Aker et~al., \emph{First direct
  neutrino-mass measurement with sub-ev sensitivity},  2021.

\bibitem{kassiopeia17}
D.~Furse, S.~Groh, N.~Trost, M.~Babutzka, J.P.~Barrett, J.~Behrens et~al.,
  \emph{Kassiopeia: a modern, extensible c++ particle tracking package},
  \href{https://doi.org/10.1088/1367-2630/aa6950}{\emph{New Journal of Physics}
  {\bfseries 19} (2017) 053012}.

\bibitem{Steinbrink2013}
N.~Steinbrink, V.~Hannen, E.L.~Martin, R.G.H.~Robertson, M.~Zacher and
  C.~Weinheimer, \emph{Neutrino mass sensitivity by {MAC-E-Filter} based
  time-of-flight spectroscopy with the example of {KATRIN}},
  \href{https://doi.org/10.1088/1367-2630/15/11/113020}{\emph{New Journal of
  Physics} {\bfseries 15} (2013) 113020}.

\bibitem{Fulst2020}
A.~Fulst, A.~Lokhov, M.~Fedkevych, N.~Steinbrink and C.~Weinheimer,
  \emph{Time-focusing time-of-flight, a new method to turn a mac-e-filter into
  a quasi-differential spectrometer},
  \href{https://doi.org/10.1140/epjc/s10052-020-08484-9}{\emph{The European
  Physical Journal C} {\bfseries 80} (2020) 956}.

\bibitem{Robertson1988}
R.G.H.~Robertson and D.A.~Knapp, \emph{Direct measurements of neutrino mass},
  \href{https://doi.org/10.1146/annurev.ns.38.120188.001153}{\emph{Annual
  Review of Nuclear and Particle Science} {\bfseries 38} (1988) 185}.

\bibitem{Iso5727}
\emph{Accuracy (trueness and precision) of measurement methods and results},
  Standard \href{https://www.iso.org/standard/11833.html}{5725-1},
  International Organization for Standardization, Geneva, CH (1994).

\bibitem{AbsCal19}
O.~Rest, D.~Winzen, S.~Bauer, R.~Berendes, J.~Meisner, T.~Thümmler et~al.,
  \emph{A novel ppm-precise absolute calibration method for precision
  high-voltage dividers},
  \href{https://doi.org/10.1088/1681-7575/ab2997}{\emph{Metrologia} {\bfseries
  56} (2019) 045007}.

\bibitem{kryptonHvCal2017}
{\scshape KATRIN} collaboration, \emph{Calibration of high voltages at the ppm
  level by the difference of \textsuperscript{83m}kr conversion electron lines
  at the {KATRIN} experiment},
  \href{https://doi.org/10.1140/epjc/s10052-018-5832-y}{\emph{The European
  Physical Journal C} {\bfseries 78} (2018) 368}.

\bibitem{KNM1Ana21}
{\scshape KATRIN} collaboration, \emph{Analysis methods for the first katrin
  neutrino-mass measurement},
  \href{https://doi.org/10.1103/PhysRevD.104.012005}{\emph{Phys. Rev. D}
  {\bfseries 104} (2021) 012005}.

\bibitem{krypton2017}
{\scshape KATRIN} collaboration, \emph{First transmission of electrons and ions
  through the {KATRIN} beamline},
  \href{https://doi.org/10.1088/1748-0221/13/04/P04020}{\emph{Journal of
  Instrumentation} {\bfseries 13} (2018) P04020}.

\bibitem{PhDKraus2016}
M.~Kraus, \emph{Energy-scale systematics at the {KATRIN} main spectrometer},
  Ph.D. thesis, Karlsruher Institut f{\"u}r Technologie (KIT), 2016.
\newblock 10.5445/IR/1000054447.

\bibitem{PhDRest2019}
O.~Rest, \emph{Precision high voltage at the {KATRIN} experiment and new
  methods for an absolute calibration at ppm-level for high-voltage dividers},
  Ph.D. thesis, {Westf{\"a}lische Wilhelms-Universit{\"a}t M{\"u}nster}, 2019.

\bibitem{Venos2018}
D.~V\'{e}nos, J.~Sentkerestiov\'{a}, O.~Dragoun, M.~Slez\'{a}k,
  M.~Ry\v{s}av\'{y} and A.~\v{S}palek, \emph{Properties of
  \textsuperscript{83m}{Kr} conversion electrons and their use in the {KATRIN}
  experiment},
  \href{https://doi.org/10.1088/1748-0221/13/02/T02012}{\emph{Journal of
  Instrumentation} {\bfseries 13} (2018) T02012}.

\bibitem{Altenmueller2020}
K.~Altenm{\"u}ller et~al., \emph{High-resolution spectroscopy of gaseous 83m kr
  conversion electrons with the katrin experiment},
  \href{https://doi.org/10.1088/1361-6471/ab8480}{\emph{Journal of Physics G:
  Nuclear and Particle Physics} {\bfseries 47} (2020) 065002}.

\bibitem{PhDFulst2020}
A.~Fulst, \emph{A Novel Quasi-Diﬀerential Method for MAC-E Filters and
  Determination and Control of the Electric Potentials of the KATRIN Experiment
  with a Stabilized Condensed Krypton Source and a UV Illumination System},
  Ph.D. thesis, {Westf{\"a}lische Wilhelms-Universit{\"a}t M{\"u}nster}, 2020.

\bibitem{allan66}
D.~Allan, \emph{Statistics of atomic frequency standards},
  \href{https://doi.org/10.1109/PROC.1966.4634}{\emph{Proceedings of the IEEE}
  {\bfseries 54} (1966) 221}.

\bibitem{KNM12019}
{\scshape KATRIN} collaboration, \emph{Improved upper limit on the neutrino
  mass from a direct kinematic method by katrin},
  \href{https://doi.org/10.1103/PhysRevLett.123.221802}{\emph{Phys. Rev. Lett.}
  {\bfseries 123} (2019) 221802}.

\end{thebibliography}\endgroup

\end{document}